\documentclass{aa}  
\usepackage{graphicx}
\usepackage{txfonts}
\usepackage{xcolor}
\usepackage[normalem]{ulem}
\usepackage[colorlinks=true,linkcolor=blue,citecolor=blue,urlcolor=blue]{hyperref}

\begin{document} 

    \title{Grand Theft Moons}
    \subtitle{Formation of habitable moons around giant planets}

    \author{Zolt\'an Dencs\inst{1,2,3},
          Vera Dobos\inst{4}
          \and
          Zsolt Reg\'aly\inst{2,3}
          }
          \institute{Gothard Astrophysical Observatory, HUN-REN - ELTE Exoplanet Research Group, Szent Imre herceg u. 112, H-9700 Szombathely, Hungary;
          \email{zdencs@gothard.hu}
          \and HUN-REN Konkoly Observatory, Research Centre for Astronomy and Earth Science, Konkoly-Thege Mikl\'os \'ut 15-17, H-1121, Budapest, Hungary
          \and CSFK, MTA Centre of Excellence, Budapest, Konkoly Thege Miklós \'ut 15-17, H-1121, Budapest, Hungary
          \and Kapteyn Astronomical Institute, University of Groningen, 9747 AD, Landleven 12, Groningen, The Netherlands\\}

    \date{Received Febr 18, 2025 / Accepted \today}

\abstract
    {Of the few thousand discovered exoplanets, a significant number orbit in the habitable zone of their star. Many of them are gas giants lacking a rocky surface and solid water reservoirs necessary for life as we know it. The search for habitable environments may extend to the moons of these giant planets. No confirmed exomoon discoveries have been made as of today, but promising candidates are known. Theories suggest that moon formation is a natural process in planetary systems.}
    {We aim to study moon formation around giant planets in a phase similar to the final assembly of planet formation. We search for conditions for forming the largest moons with the highest possibility in circumplanetary disks, and investigate whether the resulting moons can be habitable.}
    {We determined the fraction of the circumplanetary disk's mass converted into moons using numerical N-body simulations where moon embryos grow via embryo$-$satellitesimal collisions, investigated in disks around giant planets consisting of 100 fully interacting embryos and 1000 satellitesimals. In fiducial simulations, a 10\,Jupiter-mass planet orbited a solar analog star at distances of 1$-$5\,au. To determine the habitability of the synthetic moons, we calculated the stellar irradiation and tidal heating flux on these moons based on their orbital and physical parameters.}
    {The individual moon mass is found to be higher when the host planet orbits at a smaller stellar distance. However, moons leave the circumplanetary disk due to the stellar thief effect, which is stronger closer to the star. We find that 32\% of synthetic moons can be habitable in the circumstellar habitable zone. Due to the intense tidal heating, the incidence rate of moon habitability is similar at 2\,au, and decreasing to 1\% at larger distances ($<5$\,au).}
    {We conclude that the circumstellar habitable zone can be extended to moons around giant planets.} 
    

\maketitle

\section{Introduction}

The search for extraterrestrial life has traditionally meant hunting for exoplanets orbiting in the classical habitable zone of a solar analog star. However, the possibility of life should not be limited to this zone. An exomoon can also have the properties required for life as we know it (see, e.g., \citealp{Williamsetal1997,Kippingetal2009,Kaltenegger2010}). The most important properties for habitability are the solid silicate surface, which is the source of the building blocks of life, and the presence of liquid water on the surface, which acts as a solvent and a transfer medium for complex molecules \citep{Lammeretal2009}. Optimal atmospheric pressure and temperature are required for the presence of liquid water at the planetary surface. The emergence and long-term evolution of life is conceivable in an environment similar to that on Earth. However, the probability of finding biomarkers could be increased by extending the search to Earth-sized exomoons orbiting giant planets in the circumstellar habitable zone (CSHZ). Moreover, the additional heat from tidal heating may make it possible to extend the region of liquid water beyond the outer edge of the CSHZ \citep{HellerBarnes2013,DobosTurner2015}. The optimal mix of the heat from stellar irradiation, tidal heating, and other sources can provide a fertile environment for life \citep{Heller2012,Dobosetal2017}.

The existence of the moons of the Solar System's giant planets as well as planet formation theories predict that the formation of regular moons is a common process (see, e.g., \citealp{PealeCanup2015}). As of today, we know of a dozen promising exomoon candidates (e.g., \citealp{FoxWiegert2021,Ozaetal2019}), but none of them have been confirmed yet. 

Regular moons assemble simultaneously with the formation of the planetary system in a circumplanetary (CP) disk. Of the many theories of moon formation, the following three appear to be the most widely accepted to describe the origin of the moons of the Solar System. 1) In the solid-enhanced minimum mass nebula model, a subnebula is detached from the protoplanetary disk by the growing giant planet. Moon formation starts with dust coagulation in the subnebula \citep{MosqueiraEstrada2003}. Moonlets of the order of 1000\,km size can be formed in $10^3-10^4$\,years \citep{Estradaetal2009}. 2) In the gas-starve disk model, the protosatellite disk is fed by the infalling material from the circumstellar protoplanetary disk \citep{CanupWard2002,OgiharaIda2012}. The formation of Galilean moon-like satellites can even last for $10^6$\,years in this theory \citep{HellerPudritz2015}. 3) In the tidally spreading disk model, the moon formation is as if moonlets were being made on an assembly line at the Roche limit of a giant planet, while the CP disk is constantly spreading outwards from inside of the Roche limit \citep{CridaCharnoz2012}. Gas is already being dissipated from the disk, and the moonlets are being fed by the debris coming from the inside of the Roche limit \citep{Fujiietal2017}. Thus, the spreading disk describes a late formation theory.

The final assembly of regular moon formation could be characterised by collisions and pebble accretion. It can be assumed that the moons are assembled by processes similar to those of the rocky planets, but scaled down to the size of satellite systems \citep{RonnetJohansen2020}. During planet formation, planetesimals are formed from the dust component of the protoplanetary disk. In the final assembly phase, the gravitational interaction between planetesimals becomes the dominant force in the disk. Larger mass bodies accrete more mass, so they grow rapidly in the runaway growth phase \citep{Greenbergetal1978}. The perturbation of the largest mass neighbours produces an increasing velocity dispersion in the disk \citep{KokuboIda1998}. As a result, planetary embryos form in the oligarchic growth phase that are massive enough to perturb the orbits of less massive planetesimals \citep{IdaMakino1993}. Protoplanets form from the embryo$-$embryo and embryo$-$planetesimal collisions. After the gas has dissipated from the disk, planets form via protoplanet collisions in the post-oligarchic growth phase \citep{Roncoetal2015}. Based on the analogy of planet formation, in our study, we refer to planetesimals as satellitesimals and embryos as moon embryos (or simply embryos).

The formation of irregular moons (which do not form together with the planet) is usually explained by the following two theories. 1) CP disks can form after a giant impact event, final assembly takes place in these disks as well, resulting in the formation of protosatellites \citep{CanupAsphaug2001,RonnetJohansen2020}. 2) In addition, planets can acquire moons by capturing asteroids formed in the circumstellar disk \citep{DebesSigurdsson2007,Williams2013}.

In this study, we investigated the final phase of regular moon formation using numerical N-body simulations. Our model is primarily described by the final phase $-$ and its continuation $-$ of the solid-enhanced minimum mass nebula and the gas-starve CP disk scenarios. We calculated the gravitational interactions in CP disks consisting of fully interacting moon embryos embedded in a swarm of satellitesimals applying a direct N-body integrator using GPUs. The semi-major axis of the planetary orbit and the mass of the host planet were varied. The gravitational field of the star can influence moon formation by stealing moons from the CP disk. Therefore, to study this effect, we compared two fiducial models: one includes a central star, while the other does not, for which case the external perturbing source acting on the CP disk is absent. We determined the parameters necessary to form the largest number and largest mass moons.

We investigated the habitability of moons assembled in the numerical simulations using a semi-analytical code. Taking into account the stellar irradiation, the stellar light reflected by the host planet, the thermal emission of the planet, and the heating from the planet$-$moon tidal interaction, we calculated the habitability of the synthetic moons based on the properties and the orbital elements of the moons from the results of the numerical simulations. Finally, we studied the habitability of moon candidates around exoplanets.

The paper is structured as follows. Section\,\ref{sec:hab_methods} is dedicated to the method of habitability calculations and the instance of possible habitable moons around known giant exoplanets. Section\,\ref{sec:nbody_methods} presents a describe of the numerical method applied for the N-body calculations. The results of the N-body simulations and habitability calculations are presented and discussed in Sect.\,\ref{sec:results}. The main conclusions are summarized in Sect.\,\ref{sec:conclusions}.

\section{Habitability of moons}\label{sec:hab_methods}

\subsection{Heat sources of satellites}

The simplest CSHZ is a ring-shaped region around a star assuming an Earth-like planet. The extent of the ring depends on the stellar luminosity. Water can exist in liquid state on the surface of an Earth-mass ($M_{\oplus}$) planet orbiting between the boundaries of the CSHZ with an Earth-like atmosphere (e.g., \citealp{Kastingetal1993}). If the planet orbits closer to the central star than the inner edge of the CSHZ, the surface environment is too hot; if the planet orbits at a larger distance from the star than the outer edge of the CSHZ, the surface is too cold for liquid state water. However, habitable environments can be extended to the moons due to additional heat sources such as the stellar light reflected by the host planet, the thermal emission from the planet, and the heating from the planet$-$moon tidal interaction \citep{HellerBarnes2013,Dobosetal2017}. In addition to the stellar irradiation, the additional heat sources can provide a circumplanetary habitable zone (CPHZ) around the host planet for a given moon. It is important to note that the extent of the CPHZ  depends strongly on the individual parameters of the moon.

For moons, the same incoming heat levels occur at different distances from the central star than it was predicted by \cite{Kopparapuetal2014} for planets, due to the presence of additional heat sources. The global energy flux on the moons can be significantly influenced by tidal heating, which comes from the tidal energy dissipation of the planet$-$moon interactions. The tidal heating rate in a viscoelastic satellite can be calculated using the following expression (e.g., \citealp{Peale2003,MeyerWisdom2007})
\begin{equation}
    H_{\mathrm{tidal}}=\frac{21}{2}\frac{k_2}{Q}\frac{GM_{\mathrm{pl}}^2R^5ne^2}{a^6},\label{eq:tidal}
\end{equation}
where $G$ is the gravitational constant, $M_{\mathrm{pl}}$ is the planetary mass, $R$, $n$, $e$, and $a$ are the radius, mean motion, eccentricity, and semi-major axis of the satellite, assuming a synchronous rotation. The mean motion can be written as $n=(GM_{\mathrm{pl}}/a^3)^{1/2}$. $Q$ is the tidal dissipation factor and $k_2$ is the second-order Love number of the satellite. The $k_2/Q$ parameter together describes the tidal response of the body and can be replaced by the imaginary number of $k_2$, Im($k_2$) for a viscoelastic description of the rocky body \citep{Segatzetal1988}. Viscoelastic models are more realistic than models with a fixed $k_2/Q$ value because they include the temperature dependence of the viscosity ($\eta_s$) and the shear modulus ($\mu_s$) of the rocky material. For calculating Im($k_2$), we follow the Maxwell model description of \citet{Henningetal2009}, which was applied to exomoons for the first time by \citet{DobosTurner2015}. Based on \cite{HellerBarnes2013}, we also calculated the stellar irradiation, the energy flux from the stellar light reflected by the planet and the thermal emission of the planet on the surface of a moon. We used a semi-analytical method to obtain the global heat flux on the moons. For more details on the habitability calculation, see \cite{Dobosetal2017}.

\subsection{A demonstration for moon habitability}\label{sec:hab_demo}

To demonstrate that Earth-like habitable environments can also exist around giant planets, we collected a sample of giant exoplanets from the \texttt{exoplanets.org} database. We selected 461 known giant exoplanets with 1\,$M_{\mathrm{J}} < M_{\mathrm{pl}} <$ 13\,$M_{\mathrm{J}}$ ($M_\mathrm{J}$ represents Jupiter-mass). We calculated the total incoming heat flux for the selected exoplanets, assuming tidal interactions with Earth analog putative moons in $e=0.05$ orbit around their host planet. We did not specify the semi-major axes of the moon orbits. Instead, we examined the distance at which a moon must orbit the planet to receive the flux necessary for habitability. If no such distance can be defined where a putative moon can be habitable, then the planet does not have a CPHZ. When the orbit of the moon is inside the Roche limit or beyond the planetary Hill sphere, the moon also cannot be considered habitable due to the disintegration or orbital instability.

\begin{figure}
    \begin{center}                                                         
        \includegraphics[width=\hsize]{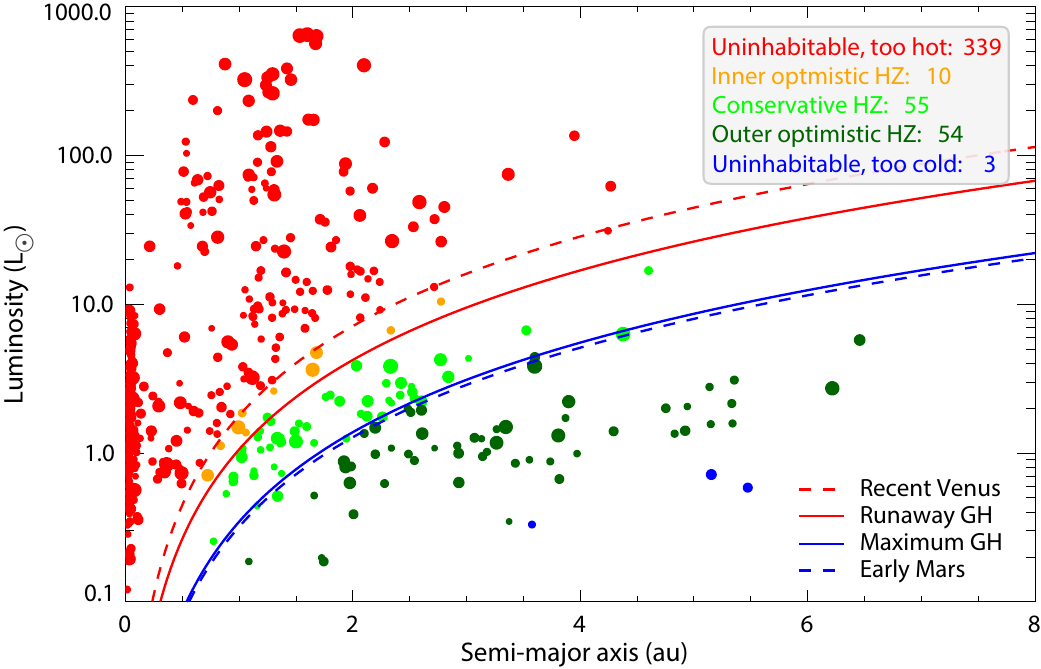}
    \end{center}
    \caption{Stellar luminosity as a function of the semi-major axis of $~1M_{\mathrm{J}} < M_{\mathrm{pl}} < 13~M_{\mathrm{J}}$ exoplanet orbits derived from \texttt{exoplanets.org}. The limits of the circumstellar habitable zone (CSHZ) are displayed with curves: recent Venus (dashed red), runaway greenhouse (solid red), maximum greenhouse (solid blue), and early Mars (dashed blue) based on \cite{Kopparapuetal2014}. Planets are represented as colored dots whose size is proportional to $M_{\mathrm{pl}}^{1/3}$. Red and blue dots represent planets that cannot host a habitable environment for Earth analog moons (they are too hot or too cold, respectively). Orange, light green and dark green colors show that the putative moon experiences enough energy to be considered habitable. The date of the \texttt{exoplanets.org} query is May 2024. We note that the purpose of the data are merely to demonstrate, using a sample of planets, that our method is valid. It is not intended to assess habitability for the most recently discovered planets.}
\label{fig:stat}
\end{figure}

Figure\,\ref{fig:stat} is an illustration of the phenomenon of how many giant exoplanets in our sample can have a putative habitable moon. The figure shows the luminosity of the exoplanets' central stars as a function of the semi-major axis of the planetary orbits. The CSHZ boundaries are calculated using Eq.\,(4) of \cite{Kopparapuetal2014}, where we assumed $S_\mathrm{eff\odot}$ values for $1~M_\oplus$ moon. Conservative and optimistic habitable zone (HZ) limits are defined based on the climate models of \cite{Kopparapuetal2013}, depending on the mass and the composition of the planetary atmospheres. The recent Venus limit refers to the inner boundary of the optimistic HZ. The runaway greenhouse limit defines the inner boundary of the conservative HZ. The environment for liquid water on the surface of an Earth-like planet orbiting in the conservative HZ is stable and optimal over long timescales. At the outer boundary of the conservative HZ, the heat from stellar irradiation drops to the level required for the maximum greenhouse limit. The outer boundary of the optimistic HZ is the early Mars limit, which assumes a relatively dense CO$_2$ atmosphere, such as that supposed to have existed on Mars~3.8~billion~years~ago. 

For moons, in addition to stellar irradiation, tidal heating can be a significant source of heat. The significance of tidal heating increases with the stellar distance. Figure\,\ref{fig:stat} shows that based on the total incoming heat flux on the moons, the host planets are classified into the following categories. $\sim$ 74\% of planets cannot host a habitable moon because the surface of a putative moon is too hot. However, about 26\% of planets can have an Earth analog habitable moon. Among these, it is important to highlight the planets orbiting outside the CSHZ. The moons around these planets can be habitable due to the additional heat provided by tidal heating. This category accounts for almost the half of the planets with a putative habitable moon in our sample. For a given stellar luminosity beyond a certain distance, even the additional heat from tidal heating is not sufficient to sustain habitability for a putative moon. We find three planets in this category.

\begin{figure}
    \begin{center}
        \includegraphics[width=\hsize]{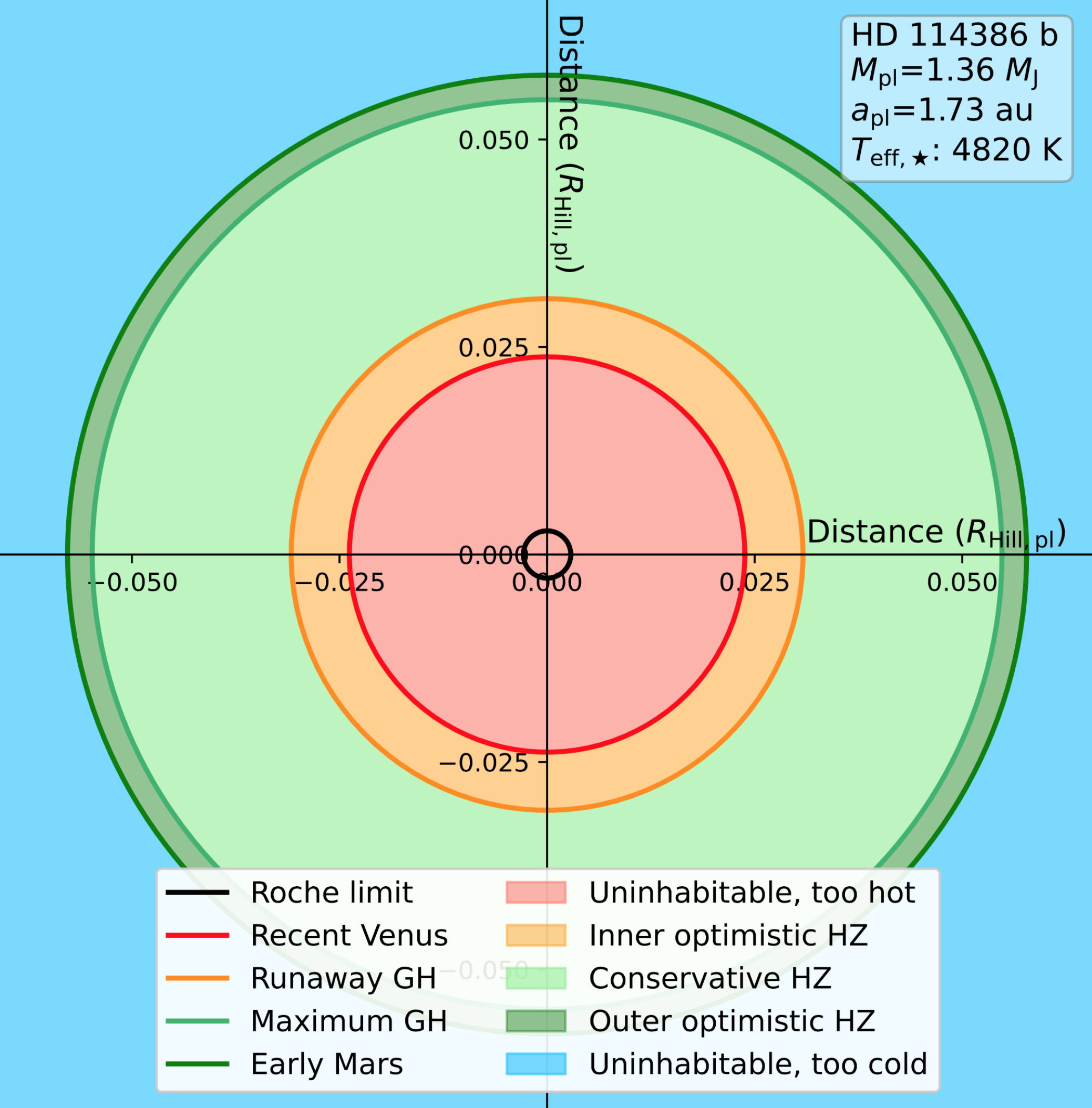}
    \end{center}
    \caption{Face-on view of the circumplanetary habitable zone of HD\,114386\,b assuming an Earth analog moon on $e=0.05$ orbit around the planet. The axes represent the distance from the host planet in the planetary Hill radius. Roche (black), recent Venus (red), runaway greenhouse (orange), maximum greenhouse (light green), and early Mars (green) limits are displayed with solid circles. The shaded zones are too hot region for habitability (red), inner optimistic habitable zone (orange), conservative habitable zone (light green), outer optimistic habitable zone (green), and too cold region for habitability (blue).}
    \label{fig:limits}
\end{figure}

Now, we estimate the region of CPHZ for a given planet at which an Earth analog moon in an $e = 0.05$ orbit can be habitable. Figure\,\ref{fig:limits} illustrates an example of the CPHZ limits from a face-on view. In the case of HD\,114386\,b, we find that a putative moon orbit within a relatively broad range, between the recent Venus and early Mars limits, can considered habitable. Since HD\,114386\,b orbits beyond the CSHZ of its K-type star, stellar irradiation gives only a relatively small contribution to the total heat flux on the moon. Instead, tidal heating becomes a crucial heat source, establishing a CPHZ where a putative moon can be habitable. Closer to the planet than the recent Venus limit, the moon would be too hot, while beyond the early Mars limit, it would be too cold for habitability.

Based on Eq.\,(\ref{eq:tidal}), one can see that the definition of the CPHZ is unique for each moon, therefore, it is difficult to define generalized distance limits for the habitable zone around planets. However, one common feature can be derived for each exoplanet: as the planet orbits farther from the star, the limits associated with the heat flux levels necessary for habitability move closer to the host planet.

\section{N-body simulations of moon formation}\label{sec:nbody_methods}

We investigate the final phase of moon formation in CP disks with numerical N-body simulations. The final assembly of rocky bodies is calculated in CP protosatellite disks, which are gas-free in our model. The disks consist of moon embryos embedded in a swarm of satellitesimals, and the only force considered in the calculation is gravity. The central star, host planet, moon embryos, and satellitesimals are calculated as gravitationally fully interacting bodies. 

The efficiency of planet formation can be used as an analogy to define the efficiency of moon formation. Moon formation efficiency can be defined as the ratio of the total mass of moons formed to the initial disk mass \citep{DencsRegaly2021}. We performed simulations to compare the formation efficiency in a stellar-centered (SC) and in a planet-centered (PC) scenario.

In the SC scenario, the central star is a Solar analog. In order to model the formation of nearly Earth-mass moons, a $10~M_{\mathrm{J}}$ giant planet is assumed to orbit the star in the fiducial model. Based on that the average density of the giant exoplanets of Fig.\,\ref{fig:stat} is found to be $\sim 1.3$\,g\,cm$^{-3}$, we assume a density of $1~\rho_{\mathrm{J}}$ and a radius of $2.61~R_{\mathrm{J}}$ for the host planet. In the additional simulations, the mass of the giant planet is set to 2 and $5~M_{\mathrm{J}}$. The semi-major axis ($a_{\mathrm{pl}}$) is set to 1, 2, 3, and 5\,au in the different simulations. The initial eccentricity and inclination of the planetary orbit are both zero. In the PC scenario, we use the same initial conditions as in the SC scenario fiducial model, but the central star is removed, the $10~M_{\mathrm{J}}$ planet is the new center of the system. In the PC simulations, there is no stellar perturbation in the system.

To determine the initial mass of the CP disk, we assume the formation of Earth-mass moons, which are important for the habitability investigations. The initial disk mass is set to $2~M_{\oplus}$, for which case at least one Earth-mass moon may form due to the fact that the formation efficiency is less than 100\%. Considering a $10~M_{\mathrm{J}}$ planet, the disk-to-planet mass ratio is found to be of the order of $10^{-4}$ in agreement with the mass ratio generally assumed for regular moon systems by \cite{CanupWard2006}. In the additional simulations with $5~M_{\mathrm{J}}$ and $2~M_{\mathrm{J}}$ host planets, the masses of the protosatellite disks are reduced by 50 and 80\%, respectively.

Assuming that a final assembly phase in a protosatellite disk is the scaled-down version that of in a protoplanetary disk, we applied similar initial numbers and properties for embryos and satellitesimals as widely used in planet assembly N-body simulations (e.g., \citealp{KokuboIda1998,Raymondetal2009,Roncoetal2015,LykawkaIto2019,Clementetal2021}). Initially, 100 embryos and 1000 satellitesimals were placed in the CP disk. We have previously shown that simulations containing fully interacting planetesimals in embryo$-$planetesimal disks are numerically convergent $-$ meaning that the initial number of embryos does not affect the results $-$ even with as few as 100 initial embryos \citep{DencsRegaly2021}.

The total embryo-to-satellitesimal mass ratio is 1:1. Thus, the individual masses of the embryos and the satellitesimals are 0.01 and $0.001~M_{\oplus}$, respectively. The density of the embryos and satellitesimals is uniformly 3\,g\,cm$^{-3}$ (see, e.g., \citealp{LykawkaIto2019}). Using a spherical approximation for the bodies, the radii of the embryos and the satellitesimals are set to $1.11\times10^{-2}~R_{\mathrm{pl}}$ ($\sim$ 2037\,km) and $5.18\times10^{-3}~R_{\mathrm{pl}}$ ($\sim$ 945\,km), respectively.

The inner and the outer edges of the disk are determined based on the Roche and the Hill radii of the host planet, respectively. In this region, the orbits of embryos and satellitesimals can be stable on long timescales. Initially, we considered the orbit of Io as the inner boundary, scaled up for the case of a $10~M_{\mathrm{J}}$ planet. Thus, the inner boundary is set at $5.16R_{\mathrm{J}}$. We note that, this is larger than the rigid Roche limit, its use significantly reduces the runtime of the simulations. In the SC scenario, the host planet's gravitational effect determines the orbit of the embryos and satellitesimals within the planetary Hill sphere. Beyond 0.49$\times$Hill radius, the embryos and satellitesimals are only weakly bound to the planet \citep{Domingosetal2006}. Therefore, we set a conservative 0.4 $\times$ Hill radius as the initial outer edge of the protosatellite disk. The Hill radius increases linearly with $a_{\mathrm{pl}}$, therefore the distance of the outer edge of the disk from the host planet increases with the stellar distance. We use an increasing disk size for 1, 2, 3, and 5\,au stellar distances. We note that, the surface number density of the embryos and satellitesimals decreases with the increasing distance from the central star. In PC scenario, we use also four different disk sizes referring to $a_{\mathrm{pl}}$=1, 2, 3, and 5\,au cases in SC scenario.

The angular positions and velocities of the embryos and satellitesimals are randomly distributed in the CP disk at the initial stage with the constraint that the initial surface number density of the disk is proportional to $R^{-1}$ \citep{AndrewsWilliams2007}. In the fiducial model, we run ten simulations for each CP disk. The initial position and velocity of the embryos and satellitesimals are redistributed ten times for statistical studies. Figure\,\ref{fig:xy} shows the face-on view of the CP disks at the four different stellar distances. The initial and the final distributions of the embryos and satellitesimals are shown in the top and bottom panels, respectively.

All simulations are calculated in three dimensions. We study dynamically cold and hot disks in order to investigate the effect of the initial eccentricity and inclination distribution of embryos and satellitesimals on the formation of moons. In the case of larger mutual orbital inclinations, collisions occur less frequently, thus, the embryo growth may be slower and the efficiency of the moon formation may be lower in hot disks than in cold disks. The initial inclination distribution of embryo and satellitesimal orbits follows Rayleigh distribution. In cold and hot disks the peaks of the inclination distributions are at $2\times10^{-4}$ and $4\times10^{-4}$, respectively. The initial eccentricity distribution of embryo and satellitesimal orbits also follows Rayleigh distribution. Based on the estimations of for example, \cite{IdaMakino1992,KokuboIda2000}, $2i \sim e$, therefore the peaks of the eccentricity distributions are at $4\times10^{-4}$ and $8\times10^{-4}$ in cold and hot disks, respectively.

The synthetic bodies interact with each other gravitationally, which can lead to the following outcomes. 1) The collision of two bodies results in one body containing the mass of the original bodies: impulse and mass conservation due to the perfectly inelastic collision. Only embryo$-$embryo and embryo$-$satellitesimal collisions are allowed. A collision occurs when the distance between the two bodies is less than the sum of their radii. 2) An object is accreted by the planet within the physical radius of the planet, but the mass of the object is not added to the mass of the planet. 3) The scattered body is accreted by the star within the stellar radius. 4) The scattered body is removed ($R$ > 1000\,au).

For the investigation, we use our own developed numerical integrator, \texttt{HIPERION}\footnote{https://www.konkoly.hu/staff/regaly/research/hiperion.html}, which is a GPU-based direct N-body integrator. The calculations were performed on NVidia Tesla K20, K40, and K80 GPUs. We apply a $4^{\mathrm{th}}$ order Hermite scheme in the simulations \citep{MakinoAarseth1992}. An adaptive shared time-step method is applied at each integration step, using the generalized $4^{\mathrm{th}}$ order Aarseth scheme \citep{PressSpergel1988,MakinoAarseth1992} with a given $\eta$ parameter, which controls the precision of the integrator. In our study, $\eta=0.025$ is found to be optimal. The total energy of the system should be conserved, therefore, the precision of the integrator can be described by the relative energy error in each iteration. The calculations are performed using double precision arithmetics. The relative energy error is determined by the instantaneous total energy ($E_{\mathrm{i}}$) and the total energy of the initial system ($E_{\mathrm{0}}$) as $(E_{\mathrm{i}}-E_{\mathrm{0}})/E_{\mathrm{0}}$ \citep{NitadoriMakino2008}. The relative energy error is found to be in the order of $10^{-10} - 10^{-8}$ with the optimal $\eta$ value at the end of the simulations.

Some embryos start to grow via embryo$-$embryo or embryo$-$satellitesimal collisions, some of them remain small in weight, others are scattered out from the protosatellite disk. The embryos that have survived the oligarchic growth phase can be considered moons at the end of the simulations. We note that on a timescale orders of magnitude longer than the interval investigated, moons can occasionally collide with each other or be accreted by the planet or the star in the post-oligarchic growth phase, however the investigation of this is beyond the scope of the recent study.

We must account for the presence of two timescales in our simulations: 1) the first one is measured in the number of the planetary orbits (this timescale is not the same in the different simulations); 2) the second one is defined by the number of orbits of the bodies in the protosatellite disk (this timescale can be unified in our fiducial model). The second timescale is more relevant in moon formation dynamics. Protoplanets form typically within $10^6-10^7$\,years timescale, which covers runaway and oligarchic growth phases (e.g., \citealp{KokuboIda2000}). It is plausible to assume that moon formation is faster than planet formation because of the spatially down scaled CP disk. The moon formation process can be completed within $10^4$\,years (e.g., \citealp{Cilibrasietal2018}), which means $\sim 10^6$ orbits in a protosatellite disk. Therefore, in our simulations, we apply an order of magnitude larger timescale: the integrations are finished after $1.5\times10^7$ orbits, counted at the inner edge of the protosatellite disk in the simulations with 10\,$M_{\mathrm{J}}$ planet. The simulations are calculated for $7.68\times10^4$, $2.71\times10^4$, $1.48\times10^4$, and $0.69\times10^4$ planetary orbits at 1, 2, 3, and 5\,au stellar distances, respectively. For $M_{\mathrm{pl}}=2~M_{\mathrm{J}}$ and $5~M_{\mathrm{J}}$, the simulations ran for a longer time because, in the case of a smaller planet mass, the inner edge of the disk is located closer to the planet, resulting in a shorter orbital period at the inner boundary of the disk. The total mass of the embryos converges to a constant level on the integration timescale.

In one specific case for each of the four orbital distances in the fiducial model, the simulations were extended for $3\times10^7$\,orbits at the inner edge (see the bottom row of Fig.\,\ref{fig:xy}) to verify the long-term stability of the resulting moons. The results confirm that no significant changes occur in the protosatellite disks and that the resulting moons' orbital elements can be considered stable over the investigated timescale.

\section{Results and discussion}\label{sec:results}

\subsection{Moon formation in the circumplanetary disk}\label{sec:results_formation}

\begin{figure*}
    \begin{center}                                                         
        \includegraphics[width=0.8\textwidth]{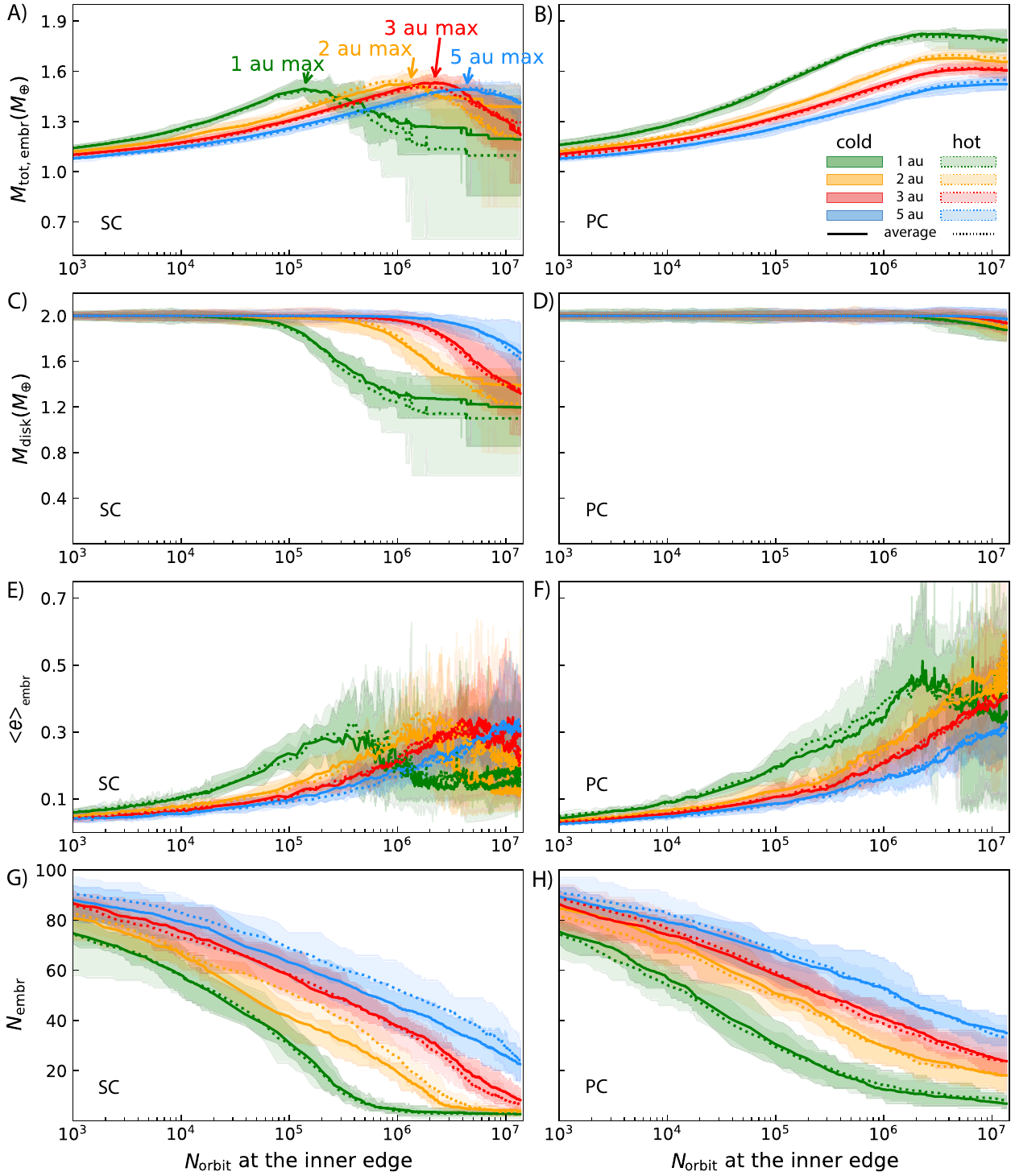}
    \end{center}
    \caption{The evolution of the moon embryos and the protosatellite disks of $10~M_{\mathrm{J}}$ host planets on a logarithmic timescale. The left and right panels show the stellar-centered (SC) and the planet-centered (PC) scenarios, respectively. Green, yellow, red, and blue colors indicate the $a_{\mathrm{pl}}$=1, 2, 3, and 5\,au simulations, respectively. Shaded regions represent the range between the minimum and maximum values from the ten simulations of each initial condition. Dark and light shadings indicate the initially cold and hot disks, respectively. Solid and dotted lines show the average values of the cold and the hot disk simulations. Panels\,A and B show the total embryo mass, panels\,C and D show the total disk mass, panels\,E and F show the average orbital eccentricity, and panels\,G and H show the number of embryos in the disks. All simulations start from the same level. Arrows indicate the maxima of the average total embryo masses in the SC scenario.}
    \label{fig:embrmass}
\end{figure*}

Panels\,A and B of Fig.\,\ref{fig:embrmass} show the evolution of the total embryo mass in the $M_{\mathrm{pl}}=10~M_{\mathrm{J}}$ SC and PC simulations, respectively. In the SC scenario, the average total embryo masses first increase, and then decrease, so they have maxima have maxima at $\sim1.58~M_{\oplus}$. In the PC scenario, the average masses also increase initially, and then only a relatively small decline occurs. The maxima of the average masses are at a higher level in the PC than in the SC simulations, and the maximum mass values are spread over a wider range in the PC than in the SC scenario. In the PC scenario, the maximum of the total embryo mass is inversely proportional with the increasing disk size.

The total embryo mass depends slightly on the stellar distance in the SC simulations. The largest mass loss occurs in the simulations where the planet orbits at 1\,au. Thus, the efficiency of moon formation is the lowest in these disks. The highest average efficiency is achieved at 2\,au stellar distance in the SC simulations in cold disks. In the PC simulations, we find that the formation efficiency is inversely proportional to the disk size in both cold and hot disks. The total mass of the embryos is larger in the PC ($1.66~M_{\oplus}$ on average) than in the SC simulations ($1.31~M_{\oplus}$ on average) at the end of the simulations. Based on this, the total embryo mass of the SC simulations could reach a higher peak, but an effect outside the disk limits the maximum mass that embryos can accrete.

\begin{figure}
    \begin{center}                                                         
        \includegraphics[width=\hsize]{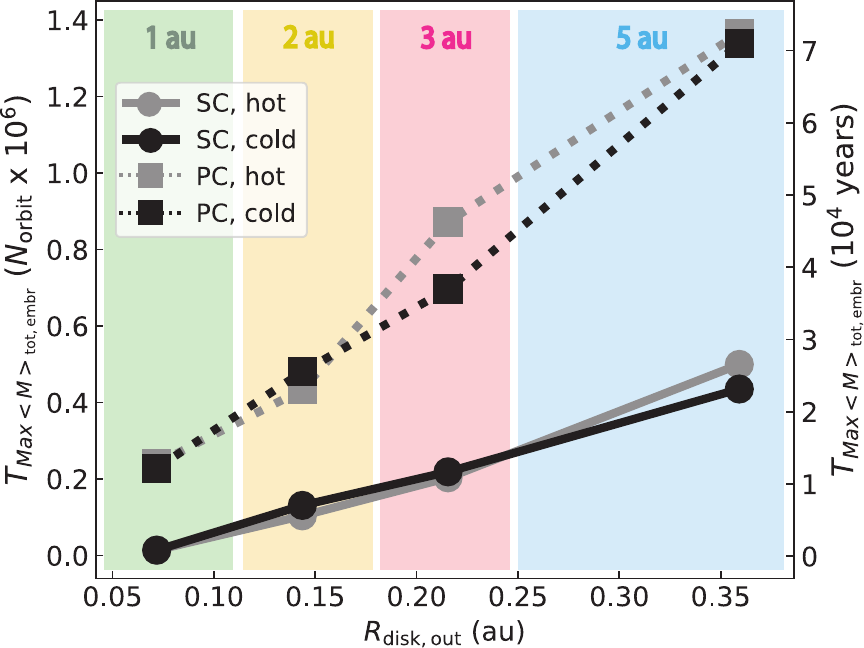}
    \end{center}
    \caption{The time for the average total embryo mass to reach its maximum as a function of the size of the disks. The time is displayed in terms of the number of orbits at the inner edge of the CP disk on the left vertical axis, as well as in years on the right vertical axis. The disk size is expressed as the distance of the disk's outer edge from the $10~M_{\mathrm{J}}$ host planet. Black and gray curves display the cold and the hot disks, respectively. Solid and dotted lines show the SC and PC scenario, respectively. The shading of the background color indicates the same stellar distances as in Fig.\,\ref{fig:embrmass}.}
\label{fig:maxtime}
\end{figure}

As an additional explanation for panels\,A and B, Fig.\,\ref{fig:maxtime} shows how the time required to reach the peak mass depends on the size of the CP disk. One can see that as the distance of the outer edge of the disk from the host planet increases, so does the time taken to reach the peak in total embryo mass in both cold and hot disks. In general, the maxima are reached later in the PC than in the SC scenario. In the $a_{\mathrm{pl}}$=1\,au SC simulations, the average total embryo mass reaches the peak after about $10^5$ orbits at the inner edge of the disk. The peak is delayed in time with the distance from the central star: in $a_{\mathrm{pl}}$=2, 3, and 5\, au simulations the total embryo mass peak is reached after $10^6$, $2\times10^6$, and $4\times10^6$ orbits, respectively. In the $a_{\mathrm{pl}}$=1\,au PC simulations, the total embryo mass reach its peak after $2\times10^6$ orbits, and the time of the peak is also delayed with increasing disk size.

After the peak level is reached in the SC scenario, the total embryo mass starts to decrease due to the embryos being scattered out from the protosatellite disk. The observed phenomenon can be explained by the following train of thought. The close encounters of the embryos excite each other's eccentricities. In smaller size disks, close encounters occur more frequently due to the higher surface density of the embryo distribution. The peak of the eccentricity distribution of the embryo orbits shifts to higher values than the initial eccentricity peak. The outermost excited embryos leave the protosatellite disk and begin to orbit the central star. Thus, in the case of SC simulations, the efficiency of moon formation is determined by two components: the surface number density of the bodies in the disk and the stealing effect of the central star.

The major trend in the embryo mass evolution is the same in the cold and hot disks. However, the standard deviation of the final total mass is larger in the hot than in the cold disks. The average efficiency of moon formation is quite similar in the cold and hot disks: the difference in the efficiency is only 2\% in favor of the cold disks. This discrepancy can be explained by the fact that the average inclination is larger in the hot than in the cold disks. The mutual inclination of the bodies in the disks is relatively large in the hot disks, thus the probability of collisions is lower in the hot disks than in the cold disks.

Panel\,C of Fig.\,\ref{fig:embrmass} displays the evolution of the total disk mass (moon embryos and satellitesimals) in SC simulations. The embryos accrete satellitesimals by collisions, which do not change the total mass of the disks. But after $7\times10^4$ orbits, the disk starts to lose mass because of some embryos and satellitesimals are removed from the disk. The total mass decay time proportional to the distance of the planet from the star. On average, 40\% and 30\% of the initial disk mass are lost in $a_{\mathrm{pl}}$=1\,au and 5\,au simulations, respectively. The mass evolution of the dynamically cold and the hot disks are very similar, however, the average disk mass is lower in the hot than in the cold disks. Panel\,D of Fig.\,\ref{fig:embrmass} shows the evolution of the disk mass in the PC scenario. In these cases, the disk mass is almost constant. Only a relatively small mass loss occurs, which can be explained mainly by the accretion of the host planet and a relatively small number of ejections from the system.

Panel\,E of Fig.\,\ref{fig:embrmass} shows the evolution of the average eccentricity of the moon embryo orbits ($\left< e \right >_\mathrm{embr}$) in the SC scenario. The evolution of $\left< e \right >_\mathrm{embr}$ correlates with the evolution of the total embryo mass for each stellar distance. The peak of the average $\left< e \right >_\mathrm{embr}$ of SC simulations is at 0.31. After reaching the peak, the average $\left< e \right >_\mathrm{embr}$ begins to decrease as the embryos in high-eccentric orbits leave the protosatellite disk due to the stellar stealing or planetary accretion. The average $\left< e \right >_\mathrm{embr}$ converges to the same level between 0.15 and 0.25. However, it has not converged by the end of the $a_{\mathrm{pl}}$=5\,au simulations.

Panel\,F of Fig.\,\ref{fig:embrmass} shows the evolution of the $\left< e \right >_\mathrm{embr}$ in the PC scenario, where the $\left< e \right >_\mathrm{embr}$ behaves the same way as in the SC scenario. This is because the high-eccentric embryos leave the disk, either by being accreted by the planet or by being ejected from the disk. However, the $\left< e \right >_\mathrm{embr}$ values are higher in the PC than in the SC simulations. This is because the high-eccentric embryos do not leave the disk in such a large numbers in the PC than in the SC scenario due to the stellar stealing. 

Panels\,G and H of Fig.\,\ref{fig:embrmass} show the evolution of the number of embryos in the SC and PC scenarios, respectively. It can be seen that the number of embryos decreases over time. In the SC scenario, the number of embryos decreases for two reasons: collisions between embryos and their ejection from the disk. Since the probability of close encounters between embryos decreases with the increasing size of the disk, the number of embryos is always higher in the larger disks. One can see that in $a_{\mathrm{pl}}$=1, 2, and 3\,au simulations, the number of embryos converges to a constant value between 1 and 25 for both cold and hot disks\footnote{Note that, Fig.\,\ref{fig:embrmass} shows the evolution on a logarithmic timescale to emphasize the maxima in total embryo mass. However, the saturation in the number of embryos is more apparent on a linear timescale. In most simulations, there is no accretion in the last 10\% of the timescale, and ejections occur only at relatively long intervals.}. The number of the surviving embryos is higher in disks more distance from the star. Note that, in the $a_{\mathrm{pl}}$=5\,au simulations, the number of embryos has not yet fully saturated. In this case, the longevity of the simulation is not sufficient for the convergence.

For an embryo to be ejected from the disk, its eccentricity must be excited above a critical eccentricity value. In the SC scenario, the embryo eccentricity must be at least 0.6 and 0.9 to be accreted by the planet from the inner and from the outer edges of the disk, respectively. To escape outward from the disk (assuming an apocenter distance of, e.g., 150\% of the disk's outer edge), only $e > 0.2$ is needed for an embryo orbiting at the outer edge, while $e > 0.5$ is required from the inner edge. This shows that embryos closer to the inner edge are more easily accreted by the planet, while those near the outer edge are more likely to be stolen by the star. In the PC scenario, there is no perturbing star affecting the disk, and the gravitational field of the planet dominates throughout the entire integration. Thus, $e > 0.6$ is also required for accretion from the inner edge, but $e \geq 1$ is required for ejection from the system. Therefore, it is harder to eject embryos from the disks in the PC scenario than in the SC scenario.

In the PC scenario, we find less frequently ejections from the disk than in the SC simulations. Therefore, the decrease in the number of embryos mainly occurs due to collisions in the PC scenario. By the end of the simulations, more embryos survive in the disks in the PC than in the SC scenario. The number of embryos converges to 5$-$40 in the PC simulations in both cold and hot disks. For the two largest initial disk sizes, the number of embryos is not yet fully saturated, as the timescale required for the perfect saturation is longer than the simulation time. Since the total embryo mass reached its maximum in all simulations, from that point on, any decay in the number of embryos can only result from embryos being ejected from the system.

The number of satellitesimals decreases for the same reason as the number of embryos. Since no collisions occur after $5\times10^6$ orbits, the number of satellitesimals only decreases due to ejection from the disks. The final number of satellitesimals is less than 300 in both cold and hot SC simulations. The mass of each satellitesimal is assumed to be unchanged, therefore their total mass is directly proportional to their number. The number of surviving satellitesimals increases with the distance from the star. 3 out of 10 $a_{\mathrm{pl}}$=1\,au simulations show no surviving satellitesimals, whereas in $a_{\mathrm{pl}}$=5\,au, at least 180 survive in all 10 simulations. In the PC scenario, the number of satellitesimals converges to 60$-$450 in both cold and hot disks. We find that a significantly larger number of satellitesimals remain in the disks in the PC scenario than in the SC scenario. This can be explained by the same critical eccentricities required for ejection as for the embryos. As a general conclusion, the effect of the star can accelerate the drop of the number of embryos and satellitesimals in the SC scenario. 

By the end of the SC simulations, the mass of the disk decreases by 30$-$40\%, approximately 20\% is accreted by the star and the planet, 10$-$20\% is ejected from the system, and 1$-$5\% remains outside the disk within the stellar-dominated field for a relatively long time. The total mass of bodies orbiting the star is proportional to the stellar distance of the former host planet, and the total mass of bodies ejected from the system is inversely proportional to that distance. By the end of the PC simulations, however, more than 90\% of the initial disk mass remains in the disks, as there is no perturbing star to steal bodies from the disks. Moreover, a significant planetary accretion occurs, which is responsible for nearly a 10\% reduction of the disk mass. The bodies ejected beyond the integration limit represent only a negligible amount of mass.

\subsection{The resulting moons}

\begin{figure}
    \begin{center}                                                         
        \includegraphics[width=\columnwidth]{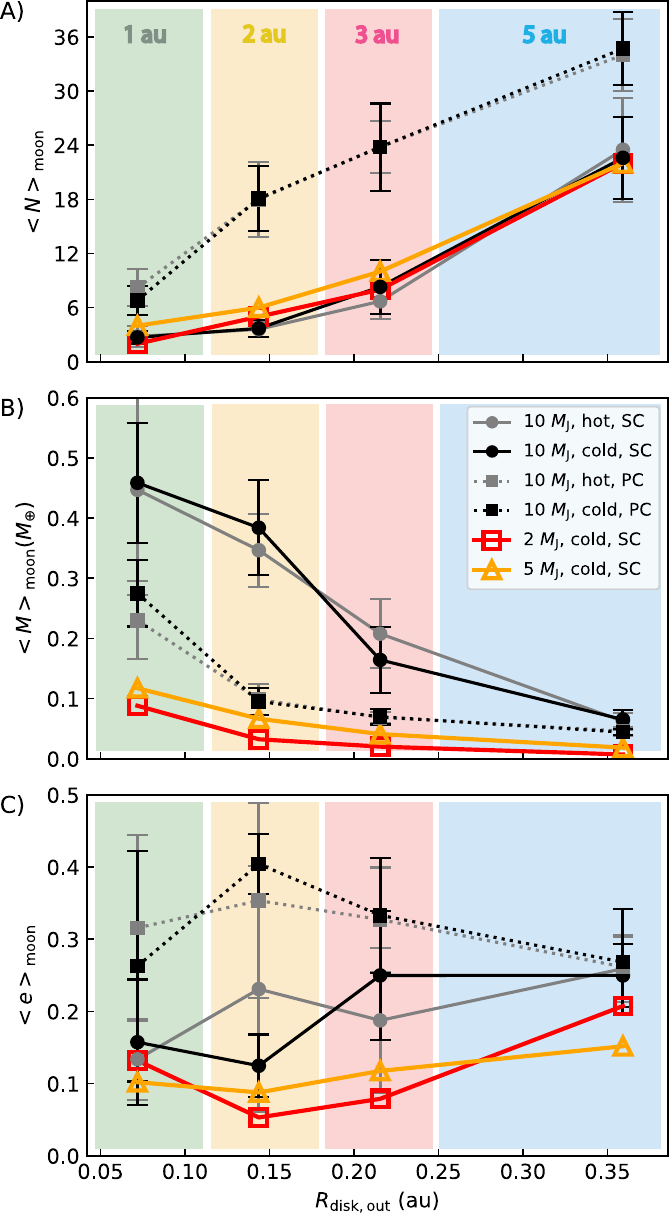}
    \end{center}
    \caption{Properties of the formed moons as a function of the size of the circumplanetary disks. Panel\,A shows the average number, panel\,B shows the average individual mass, and panel\,C shows the average eccentricity of the moons. Solid and dotted lines display the SC and the PC simulations, respectively. Black and gray lines are the simulations with cold and hot disks around $M_{\mathrm{pl}}=10~M_{\mathrm{J}}$ host planets. Vertical error bars indicate the standard deviation from the average values, which come from 10 simulations for each disk size. Red and orange lines display simulations with $M_{\mathrm{pl}}=2~M_{\mathrm{J}}$ and $5~M_{\mathrm{J}}$ host planets, respectively. The colors of the shading indicate the same stellar distances as in Fig.\,\ref{fig:embrmass}.}
    \label{fig:avgs}
\end{figure}

The surviving embryos in the disks are considered to be moons. First, we investigate the cold disk simulations. Panel\,A of Fig.\,\ref{fig:avgs} shows the average number of moons in the disks as a function of disk size. It can be seen that the average number of moons increases with disk size (thus, with stellar distance in the SC scenario). In the $M_{\mathrm{pl}}=10~M_{\mathrm{J}}$ SC simulations, the average number of moons is 2.7 and 22.6 for 1 and 5\,au stellar distances (from the average of 10 simulations), respectively. In the PC simulations, the average number of moons is 6.9 and 35 for the smallest and largest disks, respectively. This can be explained again by the stellar stealing effect in the SC cases. One can see that the average number of moons does not depend on the mass of the host planet for $M_{\mathrm{pl}}=$2, 5 and $10~M_{\mathrm{J}}$ SC simulations.

The average individual moon mass decreases with increasing disk size in each scenario, as shown in panel\,B of Fig.\,\ref{fig:avgs}. The average individual masses are $0.46~M_\oplus$ and $0.07~M_\oplus$ at 1 and 5\,au in $M_{\mathrm{pl}}=10~M_{\mathrm{J}}$ SC simulations, respectively. In the PC scenario, the average masses are $0.27~M_\oplus$ and $0.04~M_\oplus$ at 1 and 5\,au, respectively. We find that the individual moon masses are $\sim$2.2 times larger in the SC than in the PC scenario. This can be explained by the stellar perturbation of the disks and the embryo theft in the SC cases. Since there can be found, on average, $\sim$3 times fewer surviving moons in the SC than in the PC simulations, while the total moon mass is $\sim$1.2 times larger in the SC scenario than in the PC scenario. We note that the average masses of the most massive moons are $0.74~M_\oplus$ and $0.41~M_\oplus$ in the 1 and 5\,au SC simulations, respectively. Very similar to the SC values, $0.74~M_\oplus$ and $0.43~M_\oplus$ are found in the PC scenario.

One can see in panel\,B of Fig.\,\ref{fig:avgs} that the average individual moon mass increases with the mass of the host planet for each disk size. This phenomenon occurs due to the initial mass of the disks is scaled by the mass of the planet. Therefore, although the average number of moons is nearly the same in the SC simulations for disks of a given size, the average individual mass of the moons formed in lower-mass disks is smaller than in higher-mass disks.

Panel\,C of Fig.\,\ref{fig:avgs} shows the average eccentricity of the moons' orbits as a function of disk size. In general, there is no clear correlation between the average eccentricity and disk size. The average eccentricities are higher in the PC simulations ($\left< e \right >_\mathrm{moon}=0.31$) than in the SC simulations ($\left< e \right >_\mathrm{moon}=0.19$) for $M_{\mathrm{pl}}=10~M_{\mathrm{J}}$. This can be explained by that higher-eccentric embryos can survive in the disk in the PC scenario. In the $M_{\mathrm{pl}}=2~M_{\mathrm{J}}$ and $5~M_{\mathrm{J}}$ simulations, the average moon eccentricities are 0.14 and 0.15, respectively. This is because of the fact that disks around lower-mass planets initially have lower-mass embryos, which generate less eccentricity excitation in the disks\footnote{Confirmed with supplementary simulations using the parameters of the fiducial model, but with lower-mass embryos and satellitesimals.}. We emphasize that the hot disk simulations give the same results as the cold ones within the standard deviation for the average number of moons, the average individual moon mass, and the average eccentricity of moons.

\subsection{Habitability of the resulting moons}

We calculated the habitability of the resulting moons applying our semi-analytical method. We used the physical parameters and orbital elements of the resulting moons from the simulations to calculate the total heat flux reaching the surface of the moons. Based on the amount of incoming heat, we classified the moons into four categories: habitable, too hot for habitability, too cold for habitability, and too small for being habitable. A moon with a mass of $M > 1~M_{\mathrm{Mars}}$ is considered habitable if total heat flux on the moon is between the recent Venus and the early Mars flux limits. A moon with a mass smaller than $1~M_{\mathrm{Mars}}$ is not considered be able to maintain a sufficiently large atmosphere for hundreds of millions of years, which is essential for the presence of liquid water on the surface. We note that \cite{Lammeretal2014} suggest $2.5~M_{\mathrm{Mars}}$ for the minimum moon mass in the CSHZ of a Solar analog star due to the strong stellar activity that can erode the atmosphere of a low-mass moon. Although, at larger distances from the star (e.g., the distance of Mars), the impact of stellar activity on the atmosphere is weaker, thus, even a lower-mass moon can retain a sufficiently massive atmosphere.

\begin{figure*}
    \begin{center}
        \includegraphics[width=0.8\textwidth]{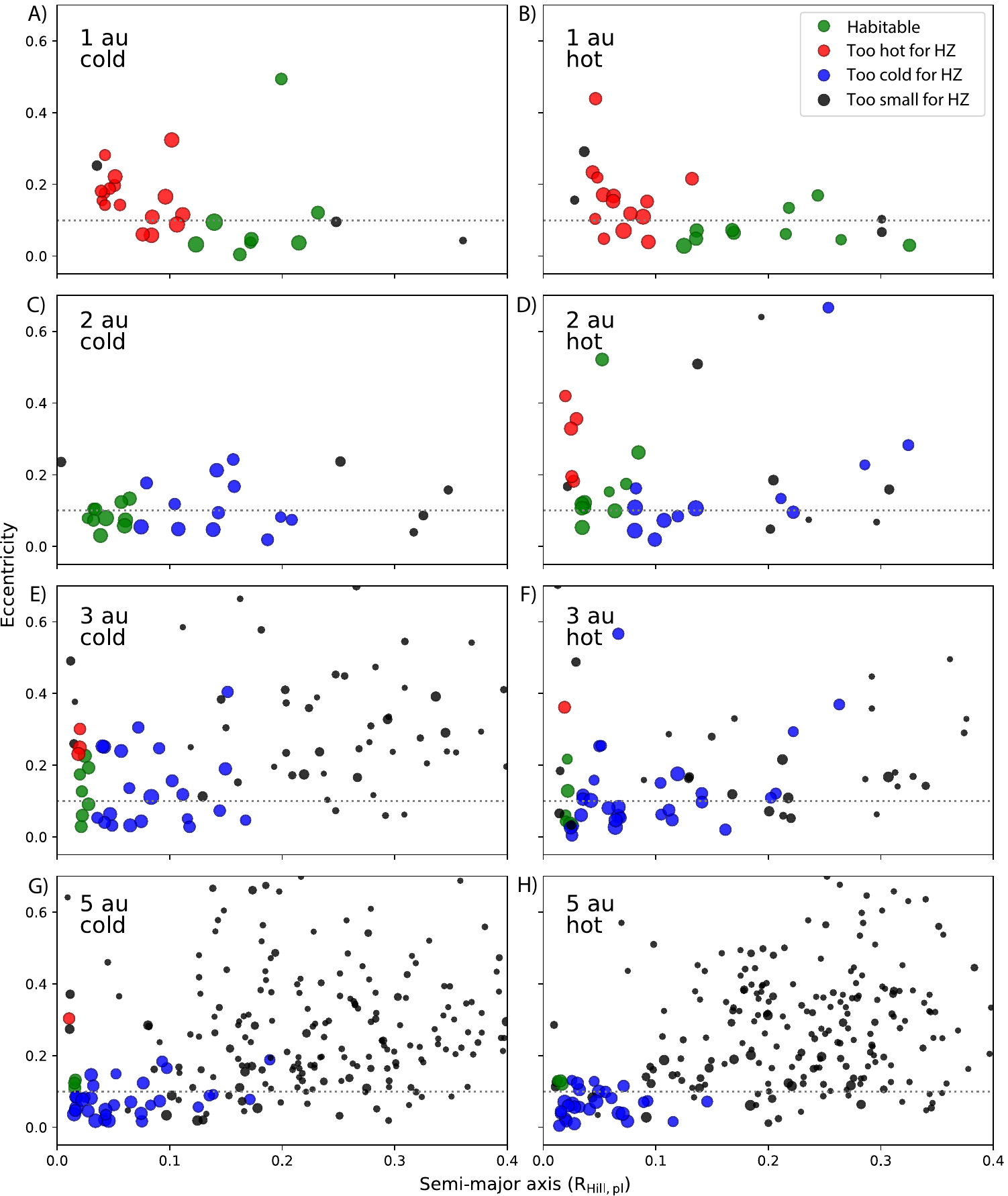}
    \end{center}
    \caption{Habitability of the moons formed in the disk of the 10\,$M_{\mathrm{J}}$ planets. Each panel shows the eccentricity as a function of the semi-major axis of the moon orbits in units of the planet's Hill radius. Left and right panels show the moons in dynamically cold and hot disks, respectively. Panels\,A, B show 1\,au, panels\,C, D show 2\,au, panels\,E, F show 3\,au, and panels\,G, H show 5\,au planetary orbit simulations. Habitable moons are indicated with green dots. Uninhabitable moons are classified into too hot (red dots), too cold (blue dots), and too small mass (black dots) categories. The size of the dots is proportional to the mass of the moons to the power of 1/3$^{\mathrm{th}}$. In each panel, the horizontal dotted line shows the $e=0.1$ upper eccentricity limit above that the confidence of the habitability calculations becomes less reliable.}
    \label{fig:habitables}
\end{figure*}

Table\,\ref{tab:hab} shows the number, the average mass, and the average eccentricity of the resulting moons of the fiducial model, separately for each habitability category. The average flux ratio of the tidal heating and stellar irradiation ($\left< F_{\mathrm{tidal}}/F_{\mathrm{stellar}} \right >$) on the surface of the moons is also displayed. To provide a generic insight about the conditions that may result in habitable moons, the results of all moon formation simulations are shown together in each category. This helps in identifying some trends. Figure\,\ref{fig:habitables} shows the habitability of the moons as functions of the eccentricities and the semi-major axes from all simulations\footnote{For better visualization, 2 $-$ 8 high-eccentric moons are missing from panels\,E, F, G, and H of Fig.\,\ref{fig:habitables}. These are too small in mass to be habitable. However, all data points are included in the statistics, see Table\,\ref{tab:hab}.}. A horizontal dotted line displays the $e=0.1$ eccentricity limit of moon orbits in each panel. Above this limit, we can estimate the global heat flux on the moons only with a relatively large uncertainty due to the characteristics of the viscoelastic model used for our calculations (see \citealp{Dobosetal2017} for more details). 

More than half of the resulting moons are too hot to be habitable around a planet orbiting at 1\,au stellar distance as can be seen in panels\,A and B of Fig.\,\ref{fig:habitables}. Moons orbiting closer than 0.11\,R$_{\mathrm{Hill, pl}}$ to the planet are too hot due to their close proximity and relatively high eccentricity. Table\,\ref{tab:hab} shows that $\left< F_{\mathrm{tidal}}/F_{\mathrm{stellar}} \right > \sim 1$ for too hot moons, which means that the tidal heating has the same significance as the stellar irradiation. However, we can find habitable moons between 0.12 and 0.33$\times$R$_{\mathrm{Hill, pl}}$ with relatively low eccentricities. At this distance from the planet, the effect of tidal heating is negligible. The average mass of the habitable moons is $0.52~M_{\oplus}$ and $0.39~M_{\oplus}$ in cold and hot disks, respectively. 

Panels\,C and D of Fig.\,\ref{fig:habitables} show the $a_{\mathrm{pl}}$=2\,au simulations. There are no moons formed in the too hot category in the cold disk simulations. However, in the hot disks, we can find a few cases. In both cold and hot disks, more than half of the moons are the too cold to be habitable. The average masses of habitable moons are found to be $0.48~M_{\oplus}$ and $0.45~M_{\oplus}$ in cold and in hot disks, respectively. Compared to the $a_{\mathrm{pl}}$=1\,au cases, the distance from the planet, where the habitable moons formed, is about 50\% shorter for $a_{\mathrm{pl}}$=2\,au cases. Here, $\left< F_{\mathrm{tidal}}/F_{\mathrm{stellar}} \right > \sim 1.3$ for both cold and hot disks, meaning that tidal heating has become the dominant heat factor for the habitability of the moons.

In the 3\,au simulations, the most populated category is the moons with $M < 1~M_{\mathrm{Mars}}$ (see panels\,E and F of Fig.\,\ref{fig:habitables}). A total of 7 and 5 habitable moons formed in the cold and in the hot disks, respectively, with average masses of $0.34~M_{\oplus}$ and $0.3~M_{\oplus}$. Here, the range in which habitable moons formed is very narrow. The significance of tidal heating is more than four times that of stellar irradiation concerning the habitability of moons. At 3\,au stellar distance, there is only a small number of moons classified as too hot for habitability.

Panels\,G and H of Fig.\,\ref{fig:habitables} show the $a_{\mathrm{pl}}$=5\,au simulations. There are an order of magnitude more moons with $M < 1~M_{\mathrm{Mars}}$ mass at $a_{\mathrm{pl}}$=5\,au than at the other stellar distances. Larger mass moons are concentrated closer to the host planet. We find that most of these higher-mass moons are too cold to be habitable. Only a few habitable moons formed at 5\,au stellar distance. Here, tidal heating flux contributes to the heating of habitable moons about 10 times more than stellar irradiation. The possibility of that a moon orbits within the narrow range where tidal heating is sufficient for habitability is very low. Therefore, we identify the less number of habitable moons in this case. The average masses of the habitable moons are $0.3~M_{\oplus}$ and $0.28~M_{\oplus}$ in the cold and hot disks, respectively. 

A general trend can be observed in the habitability of moons around $10~M_{\mathrm{J}}$ planets: as the distance from the star increases, moons orbiting closer to the planet become habitable. This is because the stellar irradiation decreases with increasing distance from the star. The significance of tidal heating in the total heat flux on the habitable moons is proportional to the stellar distance. Another trend is that the number of $M < 1~M_{\mathrm{Mars}}$ moons is proportional to the stellar distance. This can be explained by the dynamics of moon formation. We also note that for $2~M_{\mathrm{J}}$ and $5~M_{\mathrm{J}}$ planets, the average individual moon mass is smaller on average than for a $10~M_{\mathrm{J}}$ planet, thus, the number of moons that too small to be habitable is larger for smaller mass planets. Moreover, the average eccentricities are relatively lower for smaller planetary mass, reducing the effect of tidal heating. 

As result of tidal evolution, the eccentricity of moons decreases in time. Based on the Constant Phase Lag model of \cite{YoderPeale1981}, the timescale of eccentricity decay is on the order of $10^{16}$\,years for the habitable moons in Fig\,\ref{fig:habitables}. We used the following equation for the estimation
\begin{equation}
    \tau_{\mathrm{e}} = \frac{4}{63} \frac{M}{M_{\mathrm{pl}}} \left(\frac{a}{R}\right)^5 \frac{\mu_s Q}{n},\label{eq:ecc_diss}    
\end{equation}
assuming $\mu_s = 60$\,GPa and $Q = 370$ parameters which are widely used values based on Earth's average composition (see, e.g., \citealp{Jietal2010,Ferraz-Mello2013}).

We conclude that, in the Solar analog system, habitable moons are most commonly found in the $a_{\mathrm{pl}}$=1 and 2\,au simulations, regardless of whether the moons formed in a cold or in a hot CP disk. At larger distances from the star, the tidal heating become the dominant heat source for the habitability of the moons. Due to tidal heating, habitable moons can form at larger stellar distances than the outer boundary of the classical CSHZ.

\subsection{Habitability of exomoons in known systems}

Using the 461 known exoplanets from the \texttt{exoplanets.org} database presented in Sect.\,\ref{sec:hab_demo}, we selected the giant planets that most closely resemble the models calculated in our N-body simulations. We selected planets with mass $2~M_{\mathrm{J}} < M_{\mathrm{pl}} < 10~M_{\mathrm{J}}$, semi-major axis $1~\mathrm{au} < M_{\mathrm{pl}} < 5~\mathrm{au}$, eccentricity $0 < e_{\mathrm{pl}} < 0.1$, and host star luminosity $0.1~L_\odot < L_{\mathrm{star}} < 10.0~L_\odot$. We found that 9 of these planets can host a habitable moon which mass is assumed to be $1~M_\oplus$. Table\,\ref{tab:pl_data} shows the parameters of these planetary systems.

For planets orbiting beyond the outer edge of the CSHZ, tidal heating provides the habitability for the putative moon. We note that in systems with multiple planets, their gravitational interactions may destabilize the putative planet-moon system. We emphasize that our calculations assume a $1~M_\oplus$ moon.

We also investigated the habitability of 12 exomoon candidates. Table\,\ref{tab:sat_data} presents the candidates, the reason why they are not habitable, and the reference for the parameters. The eccentricity of the exomoon orbits is not available in most cases, thus, we used an estimated eccentricity of $e=0.05$ and a density of 3\,g\,cm$^{-3}$ for all exomoons. We found that the candidates are too hot to be habitable, because they are too small (Io analogs) or orbiting too close to the host planet.

\begin{table}
\centering
\fontsize{7pt}{12pt}\selectfont 
\caption{Known exoplanets that are similar to that of in our model and can host a habitable assumed 1\,$M_\oplus$ moon.}
	\begin{tabular}{c c c c c}
	\hline\hline
    Planet's & Spec. type of & Nr. of & CSHZ & Planet \\  
    name & the star & planets & relation & confirmation ref.\\
    \hline
    HD\,37605\,c & G & 2 & beyond & \cite{Wangetal2012}\\ 
    HD\,290327\,b & G & 1 & beyond & \cite{Naefetal2010}\\  
    HD\,28185\,b & G & 2 & inside & \cite{Minnitietal2009}\\ 
    47\,UMa\,b & G & 3 & beyond & \cite{GregoryFischer2010}\\
    HD\,24040\,b & G & 2 & beyond & \cite{Boisseetal2012}\\ 
    HD\,221287\,b & F & 1 & inside & \cite{Naefetal2007}\\
    HD\,10697\,b & G & 1 & inside & \cite{Wittenmyeretal2009}\\
    HD\,159868\,b & G & 2 & inside & \cite{Wittenmyeretal2012}\\
    HD\,1605\,c & K & 2 & inside & \cite{Harakawaetal2015}\\
	\hline		
	\end{tabular}
 	\label{tab:pl_data}
\end{table}

\begin{table}
\centering
\fontsize{7pt}{12pt}\selectfont 
\caption{Habitability of the exomoon candidates.}
	\begin{tabular}{c c c}
	\hline\hline
    Host planet's name & Habitability & Reference\\  
    \hline
    HD\,189733\,b & too small$^a$ & \cite{Ozaetal2019}\\ 
    KOI-268.01 & too hot & \cite{FoxWiegert2021}\\
    KOI-303.01 & too hot & \cite{FoxWiegert2021}\\
    KOI-1302.01 & too hot & \cite{FoxWiegert2021}\\
    KOI-1925.01 & too hot & \cite{FoxWiegert2021}\\
    KOI-1472.01 & too hot & \cite{FoxWiegert2021}\\
    KOI-1888.01 & too hot & \cite{FoxWiegert2021}\\
    KOI-2728.01 & too hot & \cite{FoxWiegert2021}\\
    KOI-3220.01 & too hot & \cite{FoxWiegert2021}\\
    WASP-49\,b & too small$^a$ & \cite{Ozaetal2019}\\
    WASP-76\,b & too small$^a$ & \cite{GebekOza2020}\\
    WASP-121\,b & too small$^a$ & \cite{GebekOza2020}\\
	\hline		
	\end{tabular}
    \begin{minipage}{12cm}
    \vspace{0.1cm}
    \small $^a$Io analogs 
    \end{minipage}
 	\label{tab:sat_data}
\end{table}

\subsection{Caveats of the model and future investigations}

Here, we address the important caveats of our model and possible directions for future studies. In our simulations, we applied perfectly inelastic collisions and neglected the effect of fragmentation. \cite{Chambers2013} showed that the final mass of the protoplanets does not change significantly when fragmentation is neglected and the collisions are inelastic. This means a good approximation for the runaway growth phase. However, according to \cite{HaghighipourMaindl2022}, fragmentation must be considered in the oligarchic growth model. Therefore, a future study is needed to investigate the effect of the inelastic collisions, as suggested by the model of \cite{LeinhardtStewart2012}. 

We applied the simplification that the embryos and satellitesimals have a uniform density, but the average density of the planetary or satellite seeds generally decreases with the increasing distance from the star due to the increasing mass fraction of ice in the composition (see, e.g., \citealp{Ronnetetal2017}). A more sophisticated model with distance-dependent densities should serve as the basis for a future study. 

We note that the escape of moons from a circumplanetary disk might be important for the evolution of a planetary system. We found that even $1/3~M_{\oplus}$ moons can move from the disk to circumstellar orbits. \cite{Sucerquiaetal2019} call ploonets the circumstellar objects that originate from a protosatellite disk. The orbital evolution of ploonets can be influenced by the perturbations of the parent giant planet. Several paper address this phenomenon (e.g., \citealp{Namouni2010,Yangetal2016}), however, the investigation of the long-term evolution of ploonets is beyond the scope of our study. 

Our investigation is limited to a Solar analog star. For different stellar luminosities, the limits of the circumstellar as well as circumplanetary habitable zones would show a various picture. Moreover, the planetary Hill sphere decreases with the increasing stellar mass, making stellar stealing more effective.

The formation of irregular moons also not covered here. As we see by the Solar System moons, they are captured by giant planets that orbit at larger distances than 5\,au from the Sun (see, e.g., Phoebe, Himalia). These moons orbit too far from their planet for tidal heating to be sufficient and are also too small-mass for habitability.

\section{Conclusions}\label{sec:conclusions}

In our study, we investigated the efficiency of moon formation during the final assembly phase in the circumplanetary habitable zone. We aimed to determine the conditions required for the habitability of moons. We used numerical N-body simulations with the GPU-based integrator, \texttt{HIPERION}, to compute the gravitational forces between star, planet and circumplanetary bodies. In our fiducial simulations, fully interacting moon embryos and satellitesimals were placed in a disk around a 10\,Jupiter-mass planet. The semi-major axis of the giant planet's orbit was set to 1, 2, 3, and 5\,au from a Solar analog central star. The mass of the circumplanetary disk is set up based on the canonical $10^{-4}$ satellites-to-planet mass ratio. The size of the disks increases with the stellar distance with a constant disk mass. To study the effect of the star on the moon formation, we also performed simulations without the central star. We used dynamically cold and hot disks in term of average eccentricities and inclinations. For statistical analysis we ran 10 simulations for each model, where the initial angular positions were distributed differently. In additional setup, we ran single models for 2 and 5\,Jupiter-mass planets. In the simulations, embryo$-$embryo and embryo$-$satellitesimal collisions (perfectly inelastic) are allowed, leading to the formation of moons. Using a semi-analytic code, we determined the habitability of the resulting moons after about $7\times10^4$ years. We calculated the heat flux on the surface of the moons originating from stellar irradiation, tidal interactions between the planet and the given moon, reflected light from the planet, and thermal emission of the planet. We also applied our code to determine the habitability of 12 exomoon candidates.

In the moon formation simulations, we found that about 30$-$40\% of the initial disk mass escapes from the circumplanetary disk due to the perturbations of the higher-mass embryos. Some of the embryos and satellitesimals can move beyond the outer edge of region of stable satellite orbits. Beyond the planetary Hill sphere, the embryos and satellitesimals are under the influence of the central star. Our main findings regarding the efficiency of the moon formation can be summarized as follows.

\begin{enumerate}[(i)]
	\item The number of surviving moons and the growth rate of the moon embryos depend on the surface density (thus, the size) of the protosatellite disk during the collision regime. Larger surface densities result in larger mass moons and lower numbers of the surviving moons. The number of the resulting moons increases with the stellar distance, however, their individual mass decreases. 
	\item Stellar theft imposes an upper limit on the mass and the number of moons. In fact, the efficiency of the moon formation is significantly influenced by the central star. 
    \item Due to these two factors, the highest moon formation efficiency is observed for the planet orbiting at 2\,au stellar distance.   
\end{enumerate}

With regard to the habitability of the resulting moons, our main findings are the followings.

\begin{enumerate}[(i)]
    \item Moons beyond 1\,au can become habitable only because of the tidal heating. At these stellar distances, the flux of tidal heating overcomes the stellar irradiation on the moons. 
    \item The number of the habitable moons dramatically decreases with the distance from the star beyond 2\,au. At 3\,au and 5\,au stellar distances, the circumplanetary habitable zone is extremely narrow. The optimal distance for habitability is between 1$-$2\,au stellar distances. 
    \item Although the number of moons increases with the stellar distance, the mass of these moons is too small (lighter than 1\,Mars-mass), therefore they are not habitable.
\end{enumerate}

We examined the habitability of putative Earth analog moons around 461 known giant exoplanets, selected by their mass. We found that about a quarter of these planets could have habitable environments if the habitability region is extended to the circumplanetary habitable zone. Among these giant exoplanets, we found 9 cases (Table\,\ref{tab:pl_data}) where the planet and its host star have similar properties to our model, and a 1\,Earth-mass putative moon can be habitable around the planet. Moreover, we investigated the habitability of 12 exomoon candidates (Table\,\ref{tab:sat_data}). However, none of these exomoons found to be habitable, because they are too small or too hot to be habitable. Our simulations show that moons with masses between Mars and Earth could form around planets with masses about 10 times that of Jupiter, and many of these moons could be potentially habitable at 1$-$2\,au stellar distances. These findings suggest that it is worth investigating not only rocky planets but also gas giants for Earth-like habitable environments. These locations provide suitable targets for the discovery of habitable exomoons or exomoons in general. Space telescopes offer a new opportunity for this search. From 2025, the James Webb Space Telescope (Cycle 3) will observe exoplanetary systems with exomoon candidates based on the approved proposals, for example, \cite{Casseseetal2024,Passetal2024}. The PLATO mission could provide new results for exomoon research as well after its expected launch in 2026.

\begin{acknowledgements}
    We thank the anonymous referee for useful suggestions that helped to improve the quality of the paper. This work was supported by the Hungarian Grant K119993. We acknowledge the Hungarian National Information Infrastructure Development Program (NIIF) for awarding us access to the computational resource based in Hungary at Debrecen. ZD acknowledges the support of the LP2021-9 Lend\"ulet grant of the Hungarian Academy of Sciences. ZD thanks to Gyula M. Szab\'o for the fruitful discussions.
\end{acknowledgements}

\bibliography{aanda_revised.bib}{}

\begin{thebibliography}{}
\expandafter\ifx\csname natexlab\endcsname\relax\def\natexlab#1{#1}\fi
\providecommand{\url}[1]{\href{#1}{#1}}
\providecommand{\dodoi}[1]{doi:~\href{http://doi.org/#1}{\nolinkurl{#1}}}
\providecommand{\doeprint}[1]{\href{http://ascl.net/#1}{\nolinkurl{http://ascl.net/#1}}}
\providecommand{\doarXiv}[1]{\href{https://arxiv.org/abs/#1}{\nolinkurl{https://arxiv.org/abs/#1}}}

\bibitem[{{Andrews} \& {Williams}(2007)}]{AndrewsWilliams2007}
{Andrews}, S.~M., \& {Williams}, J.~P. 2007, \apj, 659, 705,
  \dodoi{10.1086/511741}

\bibitem[{{Boisse} {et~al.}(2012){Boisse}, {Pepe}, {Perrier}, {Queloz},
  {Bonfils}, {Bouchy}, {Santos}, {Arnold}, {Beuzit}, {D{\'\i}az}, {Delfosse},
  {Eggenberger}, {Ehrenreich}, {Forveille}, {H{\'e}brard}, {Lagrange}, {Lovis},
  {Mayor}, {Moutou}, {Naef}, {Santerne}, {S{\'e}gransan}, {Sivan}, \&
  {Udry}}]{Boisseetal2012}
{Boisse}, I., {Pepe}, F., {Perrier}, C., {et~al.} 2012, \aap, 545, A55,
  \dodoi{10.1051/0004-6361/201118419}

\bibitem[{{Canup} \& {Asphaug}(2001)}]{CanupAsphaug2001}
{Canup}, R.~M., \& {Asphaug}, E. 2001, \nat, 412, 708

\bibitem[{{Canup} \& {Ward}(2002)}]{CanupWard2002}
{Canup}, R.~M., \& {Ward}, W.~R. 2002, \aj, 124, 3404, \dodoi{10.1086/344684}

\bibitem[{{Canup} \& {Ward}(2006)}]{CanupWard2006}
---. 2006, \nat, 441, 834, \dodoi{10.1038/nature04860}

\bibitem[{{Cassese} {et~al.}(2024){Cassese}, {Batygin}, {Chachan}, {Changeat},
  {Constantinou}, {Edwards}, {Kipping}, {Madhusudhan}, {Poddar}, {Teachey},
  {Tinetti}, \& {Vega}}]{Casseseetal2024}
{Cassese}, B., {Batygin}, K., {Chachan}, Y., {et~al.} 2024, {Revealing the
  Oblateness and Satellite System of an Extrasolar Jupiter Analog}, JWST
  Proposal. Cycle 3, ID. \#6491

\bibitem[{{Chambers}(2013)}]{Chambers2013}
{Chambers}, J.~E. 2013, \icarus, 224, 43, \dodoi{10.1016/j.icarus.2013.02.015}

\bibitem[{{Cilibrasi} {et~al.}(2018){Cilibrasi}, {Szul{\'a}gyi}, {Mayer},
  {Dr{\k{a}}{\.z}kowska}, {Miguel}, \& {Inderbitzi}}]{Cilibrasietal2018}
{Cilibrasi}, M., {Szul{\'a}gyi}, J., {Mayer}, L., {et~al.} 2018, \mnras, 480,
  4355, \dodoi{10.1093/mnras/sty2163}

\bibitem[{{Clement} \& {Chambers}(2021)}]{Clementetal2021}
{Clement}, M.~S., \& {Chambers}, J.~E. 2021, \aj, 162, 3,
  \dodoi{10.3847/1538-3881/abfb6c}

\bibitem[{{Crida} \& {Charnoz}(2012)}]{CridaCharnoz2012}
{Crida}, A., \& {Charnoz}, S. 2012, Science, 338, 1196,
  \dodoi{10.1126/science.1226477}

\bibitem[{{Debes} \& {Sigurdsson}(2007)}]{DebesSigurdsson2007}
{Debes}, J.~H., \& {Sigurdsson}, S. 2007, \apjl, 668, L167,
  \dodoi{10.1086/523103}

\bibitem[{{Dencs} \& {Reg{\'a}ly}(2021)}]{DencsRegaly2021}
{Dencs}, Z., \& {Reg{\'a}ly}, Z. 2021, \aap, 645, A65,
  \dodoi{10.1051/0004-6361/202039567}

\bibitem[{{Dobos} {et~al.}(2017){Dobos}, {Heller}, \& {Turner}}]{Dobosetal2017}
{Dobos}, V., {Heller}, R., \& {Turner}, E.~L. 2017, \aap, 601, A91,
  \dodoi{10.1051/0004-6361/201730541}

\bibitem[{{Dobos} \& {Turner}(2015)}]{DobosTurner2015}
{Dobos}, V., \& {Turner}, E.~L. 2015, \apj, 804, 41,
  \dodoi{10.1088/0004-637X/804/1/41}

\bibitem[{{Domingos} {et~al.}(2006){Domingos}, {Winter}, \&
  {Yokoyama}}]{Domingosetal2006}
{Domingos}, R.~C., {Winter}, O.~C., \& {Yokoyama}, T. 2006, \mnras, 373, 1227,
  \dodoi{10.1111/j.1365-2966.2006.11104.x}

\bibitem[{{Estrada} {et~al.}(2009){Estrada}, {Mosqueira}, {Lissauer},
  {D'Angelo}, \& {Cruikshank}}]{Estradaetal2009}
{Estrada}, P.~R., {Mosqueira}, I., {Lissauer}, J.~J., {D'Angelo}, G., \&
  {Cruikshank}, D.~P. 2009, in Europa, ed. R.~T. {Pappalardo}, W.~B.
  {McKinnon}, \& K.~K. {Khurana}, 27, \dodoi{10.48550/arXiv.0809.1418}

\bibitem[{{Ferraz-Mello}(2013)}]{Ferraz-Mello2013}
{Ferraz-Mello}, S. 2013, Celestial Mechanics and Dynamical Astronomy, 116, 109,
  \dodoi{10.1007/s10569-013-9482-y}

\bibitem[{{Fox} \& {Wiegert}(2021)}]{FoxWiegert2021}
{Fox}, C., \& {Wiegert}, P. 2021, \mnras, 501, 2378,
  \dodoi{10.1093/mnras/staa3743}

\bibitem[{{Fujii} {et~al.}(2017){Fujii}, {Kobayashi}, {Takahashi}, \&
  {Gressel}}]{Fujiietal2017}
{Fujii}, Y.~I., {Kobayashi}, H., {Takahashi}, S.~Z., \& {Gressel}, O. 2017,
  \aj, 153, 194, \dodoi{10.3847/1538-3881/aa647d}

\bibitem[{{Gebek} \& {Oza}(2020)}]{GebekOza2020}
{Gebek}, A., \& {Oza}, A.~V. 2020, \mnras, 497, 5271,
  \dodoi{10.1093/mnras/staa2193}

\bibitem[{{Greenberg} {et~al.}(1978){Greenberg}, {Wacker}, {Hartmann}, \&
  {Chapman}}]{Greenbergetal1978}
{Greenberg}, R., {Wacker}, J.~F., {Hartmann}, W.~K., \& {Chapman}, C.~R. 1978,
  \icarus, 35, 1, \dodoi{10.1016/0019-1035(78)90057-X}

\bibitem[{{Gregory} \& {Fischer}(2010)}]{GregoryFischer2010}
{Gregory}, P.~C., \& {Fischer}, D.~A. 2010, \mnras, 403, 731,
  \dodoi{10.1111/j.1365-2966.2009.16233.x}

\bibitem[{{Haghighipour} \& {Maindl}(2022)}]{HaghighipourMaindl2022}
{Haghighipour}, N., \& {Maindl}, T.~I. 2022, \apj, 926, 197,
  \dodoi{10.3847/1538-4357/ac4969}

\bibitem[{{Harakawa} {et~al.}(2015){Harakawa}, {Sato}, {Omiya}, {Fischer},
  {Hori}, {Ida}, {Kambe}, {Yoshida}, {Izumiura}, {Koyano}, {Nagayama},
  {Shimizu}, {Okada}, {Okita}, {Sakamoto}, \& {Yamamuro}}]{Harakawaetal2015}
{Harakawa}, H., {Sato}, B., {Omiya}, M., {et~al.} 2015, \apj, 806, 5,
  \dodoi{10.1088/0004-637X/806/1/5}

\bibitem[{{Heller}(2012)}]{Heller2012}
{Heller}, R. 2012, \aap, 545, L8, \dodoi{10.1051/0004-6361/201220003}

\bibitem[{{Heller} \& {Barnes}(2013)}]{HellerBarnes2013}
{Heller}, R., \& {Barnes}, R. 2013, Astrobiology, 13, 18,
  \dodoi{10.1089/ast.2012.0859}

\bibitem[{{Heller} \& {Pudritz}(2015)}]{HellerPudritz2015}
{Heller}, R., \& {Pudritz}, R. 2015, \apj, 806, 181,
  \dodoi{10.1088/0004-637X/806/2/181}

\bibitem[{{Henning} {et~al.}(2009){Henning}, {O'Connell}, \&
  {Sasselov}}]{Henningetal2009}
{Henning}, W.~G., {O'Connell}, R.~J., \& {Sasselov}, D.~D. 2009, \apj, 707,
  1000, \dodoi{10.1088/0004-637X/707/2/1000}

\bibitem[{{Ida} \& {Makino}(1992)}]{IdaMakino1992}
{Ida}, S., \& {Makino}, J. 1992, \icarus, 96, 107,
  \dodoi{10.1016/0019-1035(92)90008-U}

\bibitem[{{Ida} \& {Makino}(1993)}]{IdaMakino1993}
---. 1993, \icarus, 106, 210, \dodoi{10.1006/icar.1993.1167}

\bibitem[{{Ji} {et~al.}(2010){Ji}, {Sun}, {Wang}, \& {Marcotte}}]{Jietal2010}
{Ji}, S., {Sun}, S., {Wang}, Q., \& {Marcotte}, D. 2010, Journal of Geophysical
  Research (Solid Earth), 115, B06314, \dodoi{10.1029/2009JB007134}

\bibitem[{{Kaltenegger}(2010)}]{Kaltenegger2010}
{Kaltenegger}, L. 2010, \apjl, 712, L125, \dodoi{10.1088/2041-8205/712/2/L125}

\bibitem[{{Kasting} {et~al.}(1993){Kasting}, {Whitmire}, \&
  {Reynolds}}]{Kastingetal1993}
{Kasting}, J.~F., {Whitmire}, D.~P., \& {Reynolds}, R.~T. 1993, \icarus, 101,
  108, \dodoi{10.1006/icar.1993.1010}

\bibitem[{{Kipping} {et~al.}(2009){Kipping}, {Fossey}, \&
  {Campanella}}]{Kippingetal2009}
{Kipping}, D.~M., {Fossey}, S.~J., \& {Campanella}, G. 2009, \mnras, 400, 398,
  \dodoi{10.1111/j.1365-2966.2009.15472.x}

\bibitem[{{Kokubo} \& {Ida}(1998)}]{KokuboIda1998}
{Kokubo}, E., \& {Ida}, S. 1998, \icarus, 131, 171,
  \dodoi{10.1006/icar.1997.5840}

\bibitem[{{Kokubo} \& {Ida}(2000)}]{KokuboIda2000}
---. 2000, \icarus, 143, 15, \dodoi{10.1006/icar.1999.6237}

\bibitem[{{Kopparapu} {et~al.}(2014){Kopparapu}, {Ramirez}, {SchottelKotte},
  {Kasting}, {Domagal-Goldman}, \& {Eymet}}]{Kopparapuetal2014}
{Kopparapu}, R.~K., {Ramirez}, R.~M., {SchottelKotte}, J., {et~al.} 2014,
  \apjl, 787, L29, \dodoi{10.1088/2041-8205/787/2/L29}

\bibitem[{{Kopparapu} {et~al.}(2013){Kopparapu}, {Ramirez}, {Kasting}, {Eymet},
  {Robinson}, {Mahadevan}, {Terrien}, {Domagal-Goldman}, {Meadows}, \&
  {Deshpande}}]{Kopparapuetal2013}
{Kopparapu}, R.~K., {Ramirez}, R., {Kasting}, J.~F., {et~al.} 2013, \apj, 765,
  131, \dodoi{10.1088/0004-637X/765/2/131}

\bibitem[{{Lammer} {et~al.}(2009){Lammer}, {Bredeh{\"o}ft}, {Coustenis},
  {Khodachenko}, {Kaltenegger}, {Grasset}, {Prieur}, {Raulin}, {Ehrenfreund},
  {Yamauchi}, {Wahlund}, {Grie{\ss}meier}, {Stangl}, {Cockell}, {Kulikov},
  {Grenfell}, \& {Rauer}}]{Lammeretal2009}
{Lammer}, H., {Bredeh{\"o}ft}, J.~H., {Coustenis}, A., {et~al.} 2009, \aapr,
  17, 181, \dodoi{10.1007/s00159-009-0019-z}

\bibitem[{{Lammer} {et~al.}(2014){Lammer}, {Schiefer}, {Juvan}, {Odert},
  {Erkaev}, {Weber}, {Kislyakova}, {G{\"u}del}, {Kirchengast}, \&
  {Hanslmeier}}]{Lammeretal2014}
{Lammer}, H., {Schiefer}, S.-C., {Juvan}, I., {et~al.} 2014, Origins of Life
  and Evolution of the Biosphere, 44, 239, \dodoi{10.1007/s11084-014-9377-2}

\bibitem[{{Leinhardt} \& {Stewart}(2012)}]{LeinhardtStewart2012}
{Leinhardt}, Z.~M., \& {Stewart}, S.~T. 2012, \apj, 745, 79,
  \dodoi{10.1088/0004-637X/745/1/79}

\bibitem[{{Lykawka} \& {Ito}(2019)}]{LykawkaIto2019}
{Lykawka}, P.~S., \& {Ito}, T. 2019, \apj, 883, 130,
  \dodoi{10.3847/1538-4357/ab3b0a}

\bibitem[{{Makino} \& {Aarseth}(1992)}]{MakinoAarseth1992}
{Makino}, J., \& {Aarseth}, S.~J. 1992, \pasj, 44, 141

\bibitem[{{Meyer} \& {Wisdom}(2007)}]{MeyerWisdom2007}
{Meyer}, J., \& {Wisdom}, J. 2007, \icarus, 188, 535,
  \dodoi{10.1016/j.icarus.2007.03.001}

\bibitem[{{Minniti} {et~al.}(2009){Minniti}, {Butler}, {L{\'o}pez-Morales},
  {Shectman}, {Adams}, {Arriagada}, {Boss}, \& {Chambers}}]{Minnitietal2009}
{Minniti}, D., {Butler}, R.~P., {L{\'o}pez-Morales}, M., {et~al.} 2009, \apj,
  693, 1424, \dodoi{10.1088/0004-637X/693/2/1424}

\bibitem[{{Mosqueira} \& {Estrada}(2003)}]{MosqueiraEstrada2003}
{Mosqueira}, I., \& {Estrada}, P.~R. 2003, \icarus, 163, 198,
  \dodoi{10.1016/S0019-1035(03)00076-9}

\bibitem[{{Naef} {et~al.}(2007){Naef}, {Mayor}, {Benz}, {Bouchy}, {Lo Curto},
  {Lovis}, {Moutou}, {Pepe}, {Queloz}, {Santos}, \& {Udry}}]{Naefetal2007}
{Naef}, D., {Mayor}, M., {Benz}, W., {et~al.} 2007, \aap, 470, 721,
  \dodoi{10.1051/0004-6361:20077361}

\bibitem[{{Naef} {et~al.}(2010){Naef}, {Mayor}, {Lo Curto}, {Bouchy}, {Lovis},
  {Moutou}, {Benz}, {Pepe}, {Queloz}, {Santos}, {S{\'e}gransan}, {Udry},
  {Bonfils}, {Delfosse}, {Forveille}, {H{\'e}brard}, {Mordasini}, {Perrier},
  {Boisse}, \& {Sosnowska}}]{Naefetal2010}
{Naef}, D., {Mayor}, M., {Lo Curto}, G., {et~al.} 2010, \aap, 523, A15,
  \dodoi{10.1051/0004-6361/200913616}

\bibitem[{{Namouni}(2010)}]{Namouni2010}
{Namouni}, F. 2010, \apjl, 719, L145, \dodoi{10.1088/2041-8205/719/2/L145}

\bibitem[{{Nitadori} \& {Makino}(2008)}]{NitadoriMakino2008}
{Nitadori}, K., \& {Makino}, J. 2008, \na, 13, 498,
  \dodoi{10.1016/j.newast.2008.01.010}

\bibitem[{{Ogihara} \& {Ida}(2012)}]{OgiharaIda2012}
{Ogihara}, M., \& {Ida}, S. 2012, \apj, 753, 60,
  \dodoi{10.1088/0004-637X/753/1/60}

\bibitem[{{Oza} {et~al.}(2019){Oza}, {Johnson}, {Lellouch}, {Schmidt},
  {Schneider}, {Huang}, {Gamborino}, {Gebek}, {Wyttenbach}, {Demory},
  {Mordasini}, {Saxena}, {Dubois}, {Moullet}, \& {Thomas}}]{Ozaetal2019}
{Oza}, A.~V., {Johnson}, R.~E., {Lellouch}, E., {et~al.} 2019, \apj, 885, 168,
  \dodoi{10.3847/1538-4357/ab40cc}

\bibitem[{{Pass} {et~al.}(2024){Pass}, {Bean}, {Charbonneau}, {Cherubim}, \&
  {Garcia-Mejia}}]{Passetal2024}
{Pass}, E., {Bean}, J.~L., {Charbonneau}, D., {Cherubim}, C., \&
  {Garcia-Mejia}, J. 2024, {A Search for Exoplanet Satellites that are the Same
  Size as the Earth's Moon}, JWST Proposal. Cycle 3, ID. \#6193

\bibitem[{{Peale}(2003)}]{Peale2003}
{Peale}, S.~J. 2003, Celestial Mechanics and Dynamical Astronomy, 87, 129,
  \dodoi{10.1023/A:1026187917994}

\bibitem[{{Peale} \& {Canup}(2015)}]{PealeCanup2015}
{Peale}, S.~J., \& {Canup}, R.~M. 2015, in Treatise on Geophysics, ed.
  G.~{Schubert}, 559--604, \dodoi{10.1016/B978-0-444-53802-4.00177-9}

\bibitem[{{Press} \& {Spergel}(1988)}]{PressSpergel1988}
{Press}, W.~H., \& {Spergel}, D.~N. 1988, \apj, 325, 715,
  \dodoi{10.1086/166042}

\bibitem[{{Raymond} {et~al.}(2009){Raymond}, {O'Brien}, {Morbidelli}, \&
  {Kaib}}]{Raymondetal2009}
{Raymond}, S.~N., {O'Brien}, D.~P., {Morbidelli}, A., \& {Kaib}, N.~A. 2009,
  \icarus, 203, 644, \dodoi{10.1016/j.icarus.2009.05.016}

\bibitem[{{Ronco} {et~al.}(2015){Ronco}, {de El{\'\i}a}, \&
  {Guilera}}]{Roncoetal2015}
{Ronco}, M.~P., {de El{\'\i}a}, G.~C., \& {Guilera}, O.~M. 2015, \aap, 584,
  A47, \dodoi{10.1051/0004-6361/201526367}

\bibitem[{{Ronnet} \& {Johansen}(2020)}]{RonnetJohansen2020}
{Ronnet}, T., \& {Johansen}, A. 2020, \aap, 633, A93,
  \dodoi{10.1051/0004-6361/201936804}

\bibitem[{{Ronnet} {et~al.}(2017){Ronnet}, {Mousis}, \&
  {Vernazza}}]{Ronnetetal2017}
{Ronnet}, T., {Mousis}, O., \& {Vernazza}, P. 2017, \apj, 845, 92,
  \dodoi{10.3847/1538-4357/aa80e6}

\bibitem[{{Segatz} {et~al.}(1988){Segatz}, {Spohn}, {Ross}, \&
  {Schubert}}]{Segatzetal1988}
{Segatz}, M., {Spohn}, T., {Ross}, M.~N., \& {Schubert}, G. 1988, \icarus, 75,
  187, \dodoi{10.1016/0019-1035(88)90001-2}

\bibitem[{{Sucerquia} {et~al.}(2019){Sucerquia}, {Alvarado-Montes}, {Zuluaga},
  {Cuello}, \& {Giuppone}}]{Sucerquiaetal2019}
{Sucerquia}, M., {Alvarado-Montes}, J.~A., {Zuluaga}, J.~I., {Cuello}, N., \&
  {Giuppone}, C. 2019, \mnras, 489, 2313, \dodoi{10.1093/mnras/stz2110}

\bibitem[{{Wang} {et~al.}(2012){Wang}, {Wright}, {Cochran}, {Kane}, {Henry},
  {Payne}, {Endl}, {MacQueen}, {Valenti}, {Antoci}, {Dragomir}, {Matthews},
  {Howard}, {Marcy}, {Isaacson}, {Ford}, {Mahadevan}, \& {von
  Braun}}]{Wangetal2012}
{Wang}, Sharon, X., {Wright}, J.~T., {Cochran}, W., {et~al.} 2012, \apj, 761,
  46, \dodoi{10.1088/0004-637X/761/1/46}

\bibitem[{{Williams}(2013)}]{Williams2013}
{Williams}, D.~M. 2013, Astrobiology, 13, 315, \dodoi{10.1089/ast.2012.0892}

\bibitem[{{Williams} {et~al.}(1997){Williams}, {Kasting}, \&
  {Wade}}]{Williamsetal1997}
{Williams}, D.~M., {Kasting}, J.~F., \& {Wade}, R.~A. 1997, \nat, 385, 234,
  \dodoi{10.1038/385234a0}

\bibitem[{{Wittenmyer} {et~al.}(2009){Wittenmyer}, {Endl}, {Cochran},
  {Levison}, \& {Henry}}]{Wittenmyeretal2009}
{Wittenmyer}, R.~A., {Endl}, M., {Cochran}, W.~D., {Levison}, H.~F., \&
  {Henry}, G.~W. 2009, \apjs, 182, 97, \dodoi{10.1088/0067-0049/182/1/97}

\bibitem[{{Wittenmyer} {et~al.}(2012){Wittenmyer}, {Horner}, {Tuomi}, {Salter},
  {Tinney}, {Butler}, {Jones}, {O'Toole}, {Bailey}, {Carter}, {Jenkins},
  {Zhang}, {Vogt}, \& {Rivera}}]{Wittenmyeretal2012}
{Wittenmyer}, R.~A., {Horner}, J., {Tuomi}, M., {et~al.} 2012, \apj, 753, 169,
  \dodoi{10.1088/0004-637X/753/2/169}

\bibitem[{{Yang} {et~al.}(2016){Yang}, {Xie}, {Zhou}, {Liu}, \&
  {Zhang}}]{Yangetal2016}
{Yang}, M., {Xie}, J.-W., {Zhou}, J.-L., {Liu}, H.-G., \& {Zhang}, H. 2016,
  \apj, 833, 7, \dodoi{10.3847/0004-637X/833/1/7}

\bibitem[{{Yoder} \& {Peale}(1981)}]{YoderPeale1981}
{Yoder}, C.~F., \& {Peale}, S.~J. 1981, \icarus, 47, 1,
  \dodoi{10.1016/0019-1035(81)90088-9}

\end{thebibliography}
\bibliographystyle{aasjournal}

\begin{appendix}

\section{Demonstration of the change in the distributions of embryos and satellitesimals}
\label{sec:appnedixa}

\begin{figure*}
    \begin{center}                                                         
    \includegraphics[width=\textwidth]{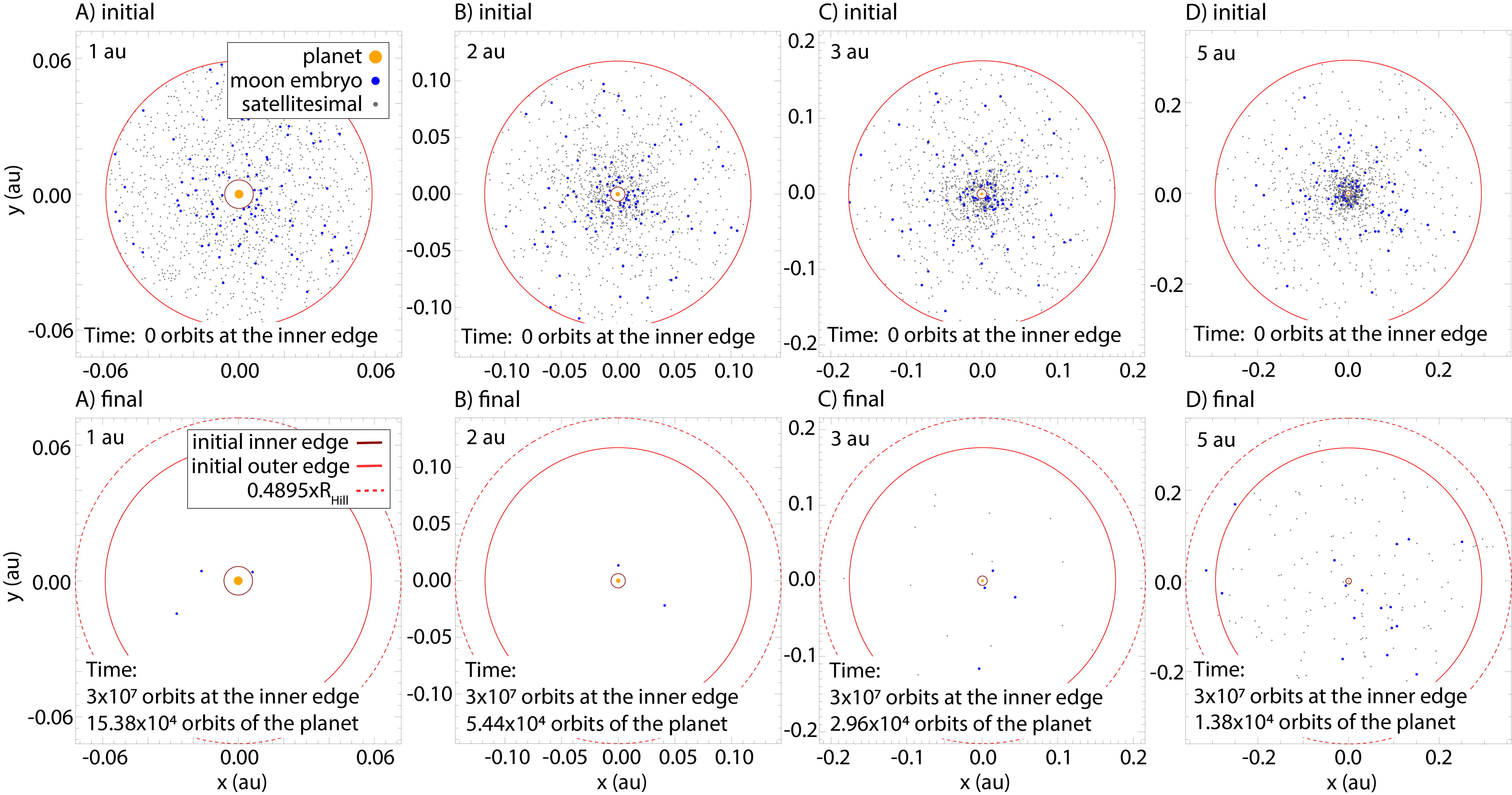}
    \caption{The distribution of the moon embryos (blue dots) and satellitesimals (gray dots) at the start (top panels) and the end (bottom panels) of the simulations in the face-on view of the protosatellite disks. Panels\,A, B, C, and D display the $a_{\mathrm{pl}}=$1, 2, 3, and 5\,au simulations. Orange dots correspond to the positions of the host planet. Brown and red circles indicate the initial inner and outer boundaries of the disks, respectively. Dashed red circle is the \cite{Domingosetal2006} suggested 0.4895$\times$Hill sphere's outer edge. The time elapsed since the start of the simulations is shown at the bottom of each panel as a number of orbits at the inner edge of the circumplanetary disk, as well as, the orbit number of the host planet.} 
    \label{fig:xy}
    \end{center}
\end{figure*}

To demonstrate the changes in the number and distribution of embryos and satellitesimals during the simulations, we provide Fig.\,\ref{fig:xy} that shows the face-on view of circumplanetary disks at four different stellar distances in the fiducial model. The top and bottom panels display the initial and final distributions of embryos and satellitesimals, respectively.

\section{Habitability classification of the resulting moons}\label{sec:appnedixb}

\begin{table*}[!h]
\centering
\fontsize{8pt}{12pt}\selectfont 
\caption{Classification of the resulting moons based on the habitability calculations.}
	\begin{tabular}{c c c c c | c c c c}
	\hline\hline
    
    & \multicolumn{4}{c}{Cold disk} & \multicolumn{4}{c}{Hot disk}\\
    & Too hot & Habitable & Too cold & Too small & Too hot & Habitable & Too cold & Too small\\
	\hline
    
    & \multicolumn{8}{c}{Distance to the star: 1\,au}\\
	\cline{2-9}
    
    Number & 16 & 8 & 0 & 3 & 14 & 10 & 0 & 4\\
    $\left< M \right > (M_{\oplus})$             & 0.426 & 0.518 & $\cdots$ & 0.034    & 0.456 & 0.392 & $\cdots$ & 0.072\\ 
    $\left< e \right >$                          & 0.163 & 0.108 & $\cdots$ & 0.13     & 0.16  & 0.073 & $\cdots$ & 0.154\\
    $\left< F_{\mathrm{tidal}} / F_{\mathrm{stellar}} \right >$  & 1.072 & 0.034 & $\cdots$ & $\cdots$ & 0.898 & 0.03  & $\cdots$ & $\cdots$\\
	\hline
        
    & \multicolumn{8}{c}{Distance to the star: 2\,au}\\
	\cline{2-9}
        
    Number & 0 & 10 & 12 & 5 & 5 & 10 & 13 & 8\\
    $\left< M \right > (M_{\oplus})$             & $\cdots$ & 0.48  & 0.352 & 0.07     & 0.353 & 0.449 & 0.356 & 0.038\\
    $\left< e \right >$                          & $\cdots$ & 0.086 & 0.111 & 0.151    & 0.296 & 0.172 & 0.21  & 0.231\\
    $\left< F_{\mathrm{tidal}} / F_{\mathrm{stellar}} \right >$  & $\cdots$ & 1.291 & 0.05  & $\cdots$ & 7.414 & 1.328 & 0.071 & $\cdots$\\
	\hline
    
    & \multicolumn{8}{c}{Distance to the star: 3\,au}\\
	\cline{2-9}
        
    Number & 3 & 7 & 21 & 52 & 1 & 5 & 29 & 32\\
    $\left< M \right > (M_{\oplus})$                     & 0.393  & 0.335 & 0.301 & 0.03     & 0.308  & 0.297 & 0.3   & 0.033\\
    $\left< e \right >$                          & 0.26   & 0.128 & 0.137 & 0.332    & 0.361  & 0.097 & 0.126 & 0.257\\
    $\left< F_{\mathrm{tidal}} / F_{\mathrm{stellar}} \right >$  & 10.188 & 4.421 & 0.369 & $\cdots$ & 12.664 & 4.516 & 0.481 & $\cdots$\\
	\hline
    
    & \multicolumn{8}{c}{Distance to the star: 5\,au}\\
	\cline{2-9}
        
    Number & 1 & 3 & 31 & 191 & 0 & 3 & 28 & 204\\
    $\left< M \right > (M_{\oplus})$                     & 0.258  & 0.303 & 0.28  & 0.018    & $\cdots$ & 0.284  & 0.274 & 0.019\\
    $\left< e \right >$                          & 0.303  & 0.122 & 0.079 & 0.287    & $\cdots$ & 0.127  & 0.063 & 0.29\\
    $\left< F_{\mathrm{tidal}} / F_{\mathrm{stellar}} \right >$  & 36.779 & 9.58  & 1.408 & $\cdots$ & $\cdots$ & 11.175 & 1.172 & $\cdots$\\
	\hline		
	\end{tabular}
 	\label{tab:hab}
\end{table*}

We provide Table\,\ref{tab:hab} to show the important properties of the resulting moons from the $a_{\mathrm{pl}}$=1, 2, 3, and 5\,au, $M_{\mathrm{pl}}=10~M_{\mathrm{J}}$ N-body simulations with dynamically cold and hot circumplanetary disks. Moons are divided into four categories based on the habitability calculations: too hot, habitable, too cold, and too small. The number, the average mass, and the average eccentricity of the moons are displayed in Table\,\ref{tab:hab}. The flux ratio of the tidal heating and stellar irradiation on the surface of the moons is given for each category. The averaged results of all moon formation simulations of the fiducial model are shown in each category.

\end{appendix}

\end{document}